\documentclass[twocolumn]{aastex7}

\usepackage{natbib}           
\usepackage{graphicx}	      
\usepackage{amsmath}	      
\usepackage{amssymb}	      
\usepackage{multirow}
\usepackage{appendix}

\usepackage{booktabs}
\usepackage{threeparttable}
\usepackage{placeins}
\usepackage{caption}

\usepackage{textcomp} 

\newcommand{\mum}{\ifmmode{\rm \mu m}\else{$\mu$m }\fi}             
\newcommand{\Msun}{\ensuremath{{\rm M}_{\odot}}}                    
\newcommand{\Msolar}{\ensuremath{{\rm M}_{\odot}}}

\newcommand{\Lsol}{L$_{\odot}$}

\providecommand{\e}[1]{\ensuremath{\times 10^{#1}}}	            
\newcommand{\chisq}{\ifmmode{\chi^{2} }\else{$\chi^2$}\fi}
\newcommand{\rchisq}{\ifmmode{\chi^{2} }\else{$\chi^2_\nu$}\fi}

\newcommand{\h}{H$_{2}$}

\received{May 14, 2025}
\accepted{September 4, 2025}

\shorttitle{First detection of ices in intermediate-mass YSOs beyond the Milky Way}   
\graphicspath{{./}{figures/}}              

\begin{document}

\title{First detection of ices in intermediate-mass young stellar objects beyond the Milky Way}

\correspondingauthor{Guido De Marchi}
\email{gdemarchi@esa.int}

\author[0000-0001-7906-3829]{Guido De Marchi}
\affil{European Space Research and Technology Centre, Keplerlaan 1, 2200 AG Noordwijk, The Netherlands}
\email{gdemarchi@esa.int}
\author[0000-0002-2667-1676]{Nolan Habel}
\affil{Jet Propulsion Laboratory, California Institute of Technology, 4800 Oak Grove Dr., Pasadena, CA 91109, USA}
\email{nolan.m.habel@jpl.nasa.gov}
\author[0000-0002-0522-3743]{Margaret Meixner}
\affil{Jet Propulsion Laboratory, California Institute of Technology, 4800 Oak Grove Dr., Pasadena, CA 91109, USA}
\email{margaret.meixner@jpl.nasa.gov}
\author[0000-0002-1892-2180]{Katia Biazzo}
\affil{INAF Osservatorio Astronomico di Roma, Via Frascati 33, 00078 Monteporzio Catone, Rome, Italy}
\email{katia.biazzo@inaf.it}
\author[0000-0002-9262-7155]{Giovanna Giardino}
\affil{ATG Europe for the European Space Agency, European Space Research and Technology Centre, Noordwijk, Netherlands}
\email{giovanna.giardino@esa.int}
\author[0000-0003-2954-7643]{Elena Sabbi}
\affil{Gemini Observatory/NSF’s NOIRLab, 950 North Cherry Avenue, Tucson, AZ, 85719, USA}
\email{elena.sabbi@noirlab.edu}
\author[0000-0001-5742-2261]{Ciaran Rogers}
\affil{Leiden Observatory, Leiden University, Leiden, the Netherlands}
\email{rogers@strw.leidenuniv.nl}
\author[0000-0002-0577-1950]{Jeroen Jaspers}
\affil{Dublin Institute for Advanced Studies, School of Cosmic Physics, Astronomy \& Astrophysics Section, Dublin 2, Ireland}
\affil{Department of Experimental Physics, Maynooth University, Maynooth, Co. Kildare, Ireland}
\email{jjaspers@cp.dias.ie}
\author[0000-0002-9573-3199]{Massimo Robberto}
\affil{Space Telescope Science Institute, 3700 San Martin Drive, Baltimore, MD 21218, USA}
\email{robberto@stsci.edu}
\author[0000-0002-6091-7924]{Peter Zeidler}
\affil{AURA for the European Space Agency, ESA Office, STScI, 3700 San Martin Drive, Baltimore, MD 21218, USA}
\email{zeidler@stsci.edu}
\author[0000-0003-4870-5547]{Olivia C. Jones}
\affil{UK Astronomy Technology Centre, Royal Observatory, Blackford Hill, Edinburgh, EH9 3HJ, UK}
\email{olivia.jones@stfc.ac.uk}
\author[0000-0003-2902-8608]{Katja Fahrion}
\affil{Department of Astrophysics, University of Vienna, T\"{u}rkenschanzstra{\ss}e 17, 1180 Wien, Austria}
\email{katja.fahrion@univie.ac.at}
\author[0000-0002-2954-8622]{Alec S. Hirschauer}
\affil{Department of Physics \& Engineering Physics, Morgan State University, 1700 East Cold Spring Lane, Baltimore, MD 21251, USA}
\email{alechirschauer@gmail.com}
\author[0000-0002-4834-369X]{Charles D. Keyes}
\affil{Space Telescope Science Institute, 3700 San Martin Drive, Baltimore, MD 21218, USA}
\email{keyes@stsci.edu}
\author[0000-0002-0322-8161]{David R. Soderblom}
\affiliation{Space Telescope Science Institute, 3700 San Martin Drive, Baltimore, MD 21218, USA}
\email{drs@stsci.edu}
\author[0000-0003-4023-8657]{Laura Lenki\'{c}}
\affil{IPAC, California Institute of Technology, 1200 East California Boulevard, Pasadena, CA 91125, USA}
\email{laura.lenkic@gmail.com}
\author[0000-0001-6576-6339]{Omnarayani Nayak}
\affil{Goddard Space Flight Center, 8800 Greenbelt Road, Greenbelt, MD 20771, USA}
\email{omnarayani.nayak@nasa.gov}
\author[0000-0001-9737-169X]{Bernhard Brandl}
\affil{Leiden Observatory, Leiden University, Leiden, the Netherlands}
\email{brandl@strw.leidenuniv.nl}

\begin{abstract}
Using NIRSpec on JWST, we studied a sample of 15 intermediate-mass ($1.8-4.1$\,\Msolar) young stellar objects (YSOs) previously identified with MIRI photometry in the low-metallicity NGC\,346 star-forming cluster in the Small Magellanic Cloud (SMC). All objects, observed in the $1.7-5.3$\,{\textmu}m range, show strong hydrogen recombination lines in the Paschen, Brackett, Pfund, and Humphreys series, confirming their very young ages. The spectra of 11 YSOs show prominent absorption bands from the three most important ice species (H$_2$O, CO$_2$, CO), marking the first detection of these ices in intermediate-mass YSOs beyond our Galaxy. In three YSOs, water ice appears to be in crystalline form. In some objects, we also detect $^{13}$CO$_2$ and OCS ices --- never before observed beyond the Milky Way (MW) --- and methanol ice in at least one star. We compared the column densities of H$_2$O, CO$_2$, and CO ices with those measured in more and less massive protostars in the MW and Large Magellanic Cloud, finding that in NGC 346 ice column densities reach values nearly an order of magnitude lower than in more massive objects ($\sim1\times10^{17}$\,cm$^{-2}$ for water and $\sim1\times10^{16}$\,cm$^{-2}$ for CO$_2$ and CO). However, the relative proportions of the ice species abundances do not differ from those in massive MW YSOs. This suggests that metallicity may not significantly affect ice chemistry in protoplanetary discs and that, shielded by the protostellar envelope or deep in the midplane, circumstellar material is likely impervious to the radiation environment.

\end{abstract}


\section{Introduction} 
\label{sec:intro}

The circumstellar environments of young stellar objects (YSOs) have rich organic chemistry that lays the ground work for planetary system formation. In particular, YSOs have notable ice features of H$_2$O, CO, CO$_2$, and more complex species (CH$_3$OH, CH$_3$OCHO; \citealt{nayaketal2024, pontoppidanetal2008}) that are one of the sources of complex organic molecules thought to be the building blocks for life \citep{ehrenfreundcharnley2000, fedoseevetal2017}. The three most abundant astronomical ices are H$_2$O, CO, and CO$_2$ \citep[e.g.][]{vandishoeck2004} and most measurements of ices are relative to H$_2$O, which is typically the most prevalent. 
 
However, our current understanding of ice chemistry is based on predominantly Milky Way (MW) sources at solar metallicity. The Large Magellanic Cloud (LMC) with $Z \simeq 1/3 Z_\odot$ \citep{hilletal1995} and the Small Magellanic Cloud (SMC) with $1/8 Z_\odot < Z <  1/5 Z_\odot$  \citep{russelldopita1992, rollestonetal1999, leeetal2005, perezmonterodiaz2005} provide lower-metallicity environments comparable to those in place at the peak epoch of star formation in the Universe at redshift $\sim 2$ \citep{peietal1999}. Furthermore, in the LMC and SMC gas-to-dust mass ratios are, respectively, 190--565 and 480--2100 according to depletion measurements \citep{tchernyshyovetal2015}, which in turn are 2 and 10 times higher than in the Galaxy. The lower dust content in both galaxies increases the ambient UV interstellar radiation field to levels higher than in the solar neighborhood, by approximately a factor of $2.1$ in the LMC \citep{bernardetal2008}, and an estimated $4-10$ times in the SMC \citep{vangioniflametal1980, pradhanetal2011}, possibly also mimicking the radiation environment typical of younger star-forming galaxies. 

Because of their proximity and well known distances, the LMC (52 kpc, \citealt{panagia1999}) and SMC (61 kpc, \citealt{kellerwood2006}) have been the targets of a number of detailed studies of ices in YSOs \citep{oliveiraetal2009, shimonishietal2010, oliveiraetal2011, oliveiraetal2013}, but so far these studies were necessarily limited to high-luminosity YSOs with masses above 8\,\Msolar {(in the following we will refer to these sources as ``high-mass YSOs'')}. Thanks to the high sensitivity and resolving power of JWST, it is now possible to probe the chemistry of discs also around smaller protostars.  

The ``Surveying the Agents of Galaxy Evolution'' (SAGE) observing programmes with the Spitzer Space Telescope revealed thousands of massive YSOs in the LMC \citep{whitneyetal2008, gruendlchu2009} and SMC \citep{sewiloetal2013}. Follow-up spectroscopy with Spitzer IRS revealed 53 YSOs with prominent CO$_2$ ice features at $15.2$\,{\textmu}m \citep{oliveiraetal2011, oliveiraetal2009, sealeetal2011}. Many of the CO$_2$ ice features in the LMC are double-peaked, indicating thermal annealing (heating of the ices that causes crystallization that changes the features). \cite{sealeetal2011} found that in the LMC the majority of the CO$_2$ ice is locked in polar ice matrices, but the fraction of polar ice matrices is lower than in the MW. In addition, the methanol (CH$_3$OH) content of the ice mixtures appears lower in the LMC sources with CO$_2$ ices that are either best fit with laboratory ice mixtures composed of less than 50\,\% CH$_3$OH, or with no CH$_3$OH. 

The studies of high-mass YSOs \citep{gerakinesetal1999, oliveiraetal2009, shimonishietal2010, sealeetal2011} have revealed a systematic difference between the LMC and MW in the ratio of the three major ice abundances. \cite{oliveiraetal2011} found that the CO$_2$ ice is more abundant in the LMC YSOs compared to MW YSOs \citep{gerakinesetal1999} with an average CO$_2$/H$_2$O column density ratio of $0.33$ as opposed to $\sim 0.2$. \cite{oliveiraetal2011, oliveiraetal2013} suggested that these larger ratios resulted from a reduced water column density in the envelopes of LMC YSOs, due to the stronger UV radiation field and to the reduced dust-to-gas ratio at lower metallicity. However, this trend does not seem to extend to the lower-metallicity SMC ($Z\simeq 1/8 Z_\odot$), where observations of massive YSOs {by \cite{oliveiraetal2011}} revealed a CO$_2$/H$_2$O column density ratio in line with the one in the MW. Furthermore, while \cite{oliveiraetal2011} found water, CO$_2$, and CO ices in all LMC YSOs, they did not detect any CO ice in the SMC objects.

In this work, we present JWST/NIRSpec observations of 15 YSOs in the SMC star-forming cluster NGC\,346 {in the range $1.7-5.3$\,{\textmu}m} and identify for the first time the clear presence of CO ices in this galaxy. The structure of the paper is as follows. In Section 2 we discuss how the YSOs in our sample were selected. Section 3 addresses the data reduction procedure and spectroscopic analysis. The ice absorption bands and the column density measurements are presented in Section 4, and discussed in Section 5, together with our conclusions. 

\section{Source Selection}

The observations presented here were obtained in the context of JWST’s guaranteed time observing (GTO) program 1227, which was structured in several parts and epochs. The initial epoch of observation contained both a spectroscopic and an imaging component. The NIRSpec instrument was used to observe pre-main sequence (PMS) sources identified previously by \cite{demarchietal2011} from the HST imaging of \cite{sabbietal2007}. The resulting spectra are presented and discussed in \cite{demarchietal2024}. At the same time, the NIRCam and MIRI instruments were used to image NGC\,346 in a total of eleven filters, covering the range from 1 to 21\,{\textmu}m, to study the stellar and pre-stellar populations  

\begin{deluxetable*}{llcccccccc}[t]
\tiny
\tablecaption{Properties of our candidate YSOs as derived from SED fitting. \label{tab1}}
\tablehead{ \colhead{ID} &\colhead{Habel Number} & \colhead{RA} & \colhead{Dec} & \colhead{$m_{277}$} & \colhead{Radius} & \colhead{Temperature} & \colhead{Luminosity}  & \colhead{Mass} & \colhead{$A_V$} \\
\colhead{} & \colhead{} & \colhead{} & \colhead{} & \colhead{} & \colhead{[$\mathrm{R_{\odot}}$]} &  \colhead{[K]} & \colhead{[$\mathrm{L_{\odot}}$]} & \colhead{[$\mathrm{M_{\odot}}$]} & \colhead{} }
\decimals
\startdata
\hline
90013~~~~~~ & CN92877 & 14.771638870 & -72.18894958 & ~~17.1~~~ & 9.2&  6610& 145.0& 4.1&  ~~3.4~~ \\
90028 & CN28745 & 14.701727867 & -72.17443848 & 18.5 & 8.7&  6655& 132.9& 4.0&  9.3 \\
90038 & CN93334 & 14.766669273 & -72.18464661 & 19.0 & 4.1&  7424&  45.4& 3.0& 18.8 \\
90052 & CN95729 & 14.834618568 & -72.18880463 & 18.8 & 7.9&  5582&  54.0& 3.1& 11.7 \\
90058 & CN44380 & 14.726758003 & -72.19560242 & 19.6 & 2.7&  6772&  14.1& 2.1& 13.7 \\
90063 & CN126460& 14.753812790 & -72.18405914 & 18.8 & 7.1&  5854&  53.9& 3.1& 10.9 \\
90069 & CN50347 & 14.828221321 & -72.15517426 & 19.0 & 1.3&  8924&   9.9& 1.9&  1.4 \\
90070 & CN50637 & 14.823440552 & -72.15420532 & 19.9 & 2.0&  7938&  14.9& 2.2&  9.3 \\
90073 & CN50580 & 14.824531555 & -72.15473175 & 19.7 & 3.3&  8302&  45.7& 3.0& 12.1 \\
90077 & CN56020 & 14.800792694 & -72.16074371 & 20.3 & 1.5& 11180&  31.0& 2.7&  --- \\
90091 & CN21701 & 14.727843285 & -72.16746521 & 19.5 & 4.5&  6528&  33.4& 2.7&  8.1 \\
90092 & CN126153& 14.756627083 & -72.18453979 & 20.2 & 4.2&  5795&  17.5& 2.3&  8.6 \\
90096 & CN20310 & 14.745772362 & -72.16602325 & 20.3 & 2.0& 11720&  71.1& 3.4&  8.1 \\
90108 & CN192819& 14.836621284 & -72.14379120 & 21.3 & 2.7&  5889&   7.8& 1.8& 27.3 \\
90123 & CN188170& 14.755599022 & -72.16664124 & 22.4 & 2.2&  6360&   7.4& 1.8& 30.0 \\
\hline
\enddata
\tablecomments{The columns are as follows: (1) ID of the source as per the NIRSpec proposal; (2) arbitrary catalog number of sources appearing in \cite{habeletal2024}; (3) and (4) J2000.0 coordinates; (5) magnitude in the NIRCam F227W band; (6) best-fit radius of YSO candidate; (7) best-fit temperature of YSO candidate; (8) luminosity of YSO candidate calculated from the best-fit radius and best-fit temperature using the relationship {$L=4 \pi r^2 \sigma T^4$}; (9) mass of the YSO candidate calculated using $L \propto M^{3.5}$; (10) estimated extinction in the $V$ band. }
\end{deluxetable*}

\noindent (\citealt{jonesetal2023, habeletal2024}, Jaspers et al. in prep.) and to serve as a form of pre-imaging. The second epoch was reserved for follow-up spectroscopic observations of embedded, lower-mass YSOs identified and confirmed by photometric analysis of the near- and mid-IR imaging from the first epoch. The available time was divided between observations with MIRI's Medium Resolution Spectroscopy (MRS) mode to target five intermediate to high-mass YSOs (Habel et al. submitted) and the NIRSpec Micro Shutter Assembly (MSA) observations described here.

\begin{figure*}[h]
\centering
\resizebox{\hsize}{!}{
\includegraphics{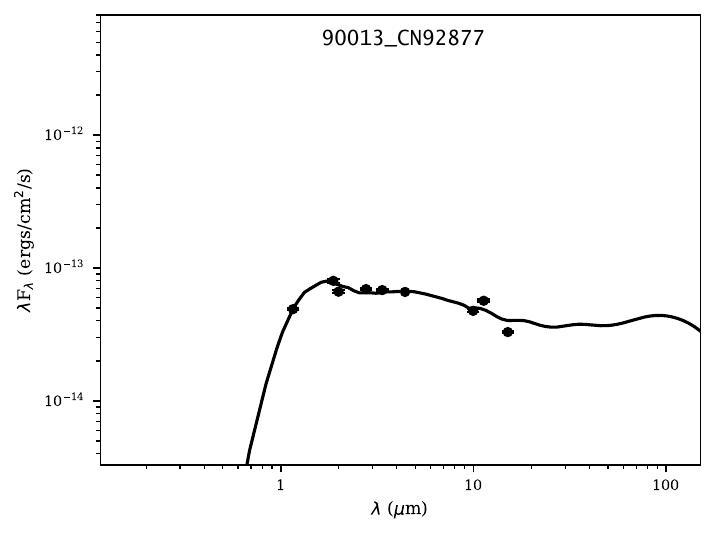}
\includegraphics{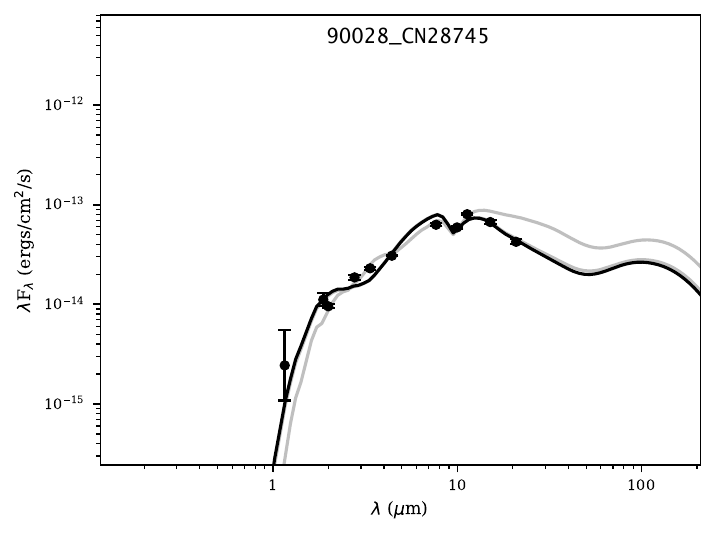}
\includegraphics{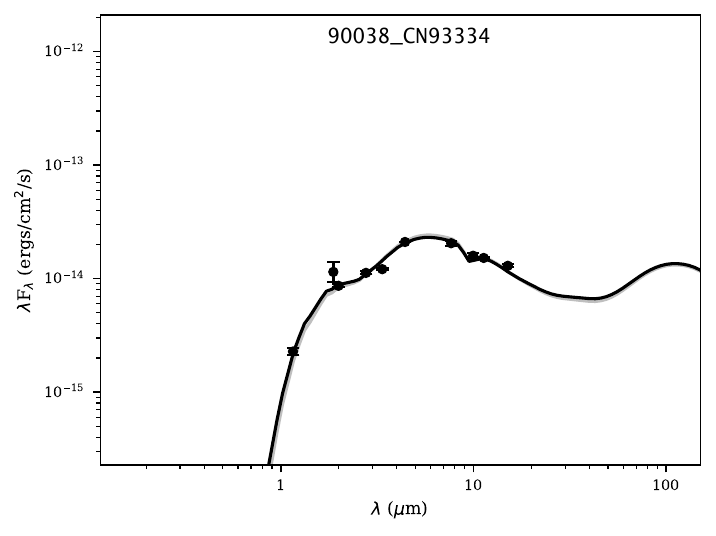}}
\resizebox{\hsize}{!}{
\includegraphics{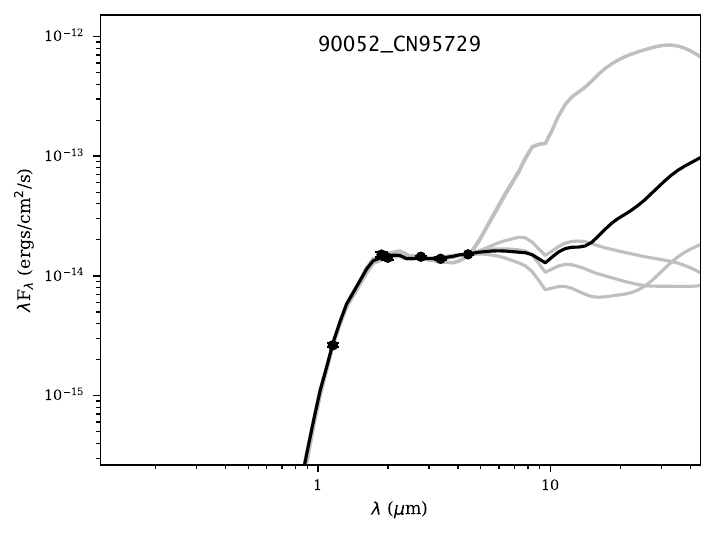}
\includegraphics{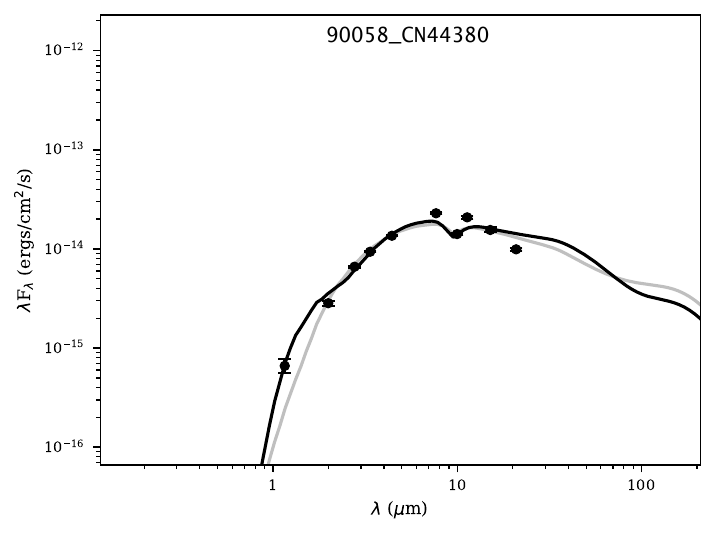}
\includegraphics{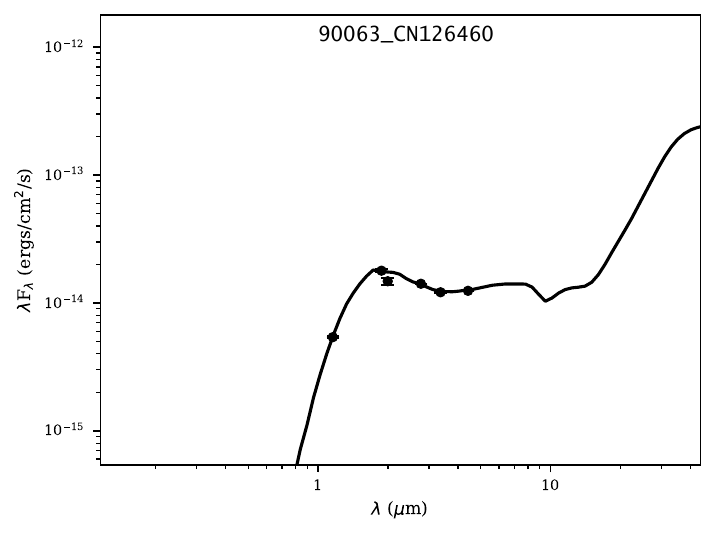}}
\resizebox{\hsize}{!}{
\includegraphics{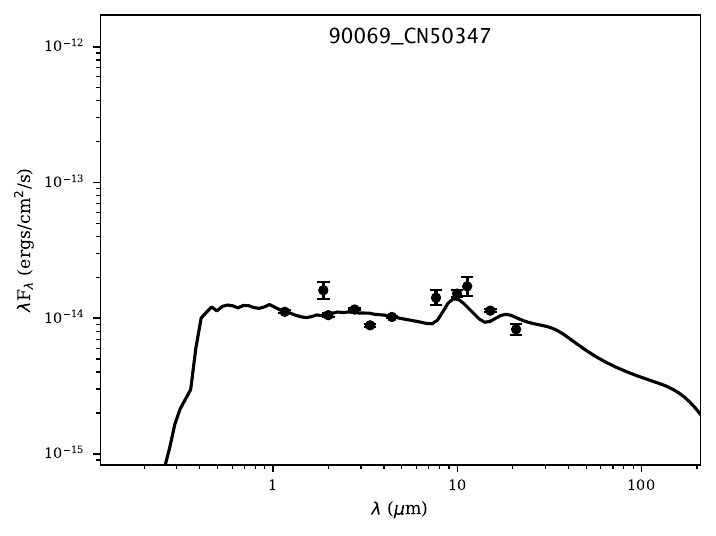}
\includegraphics{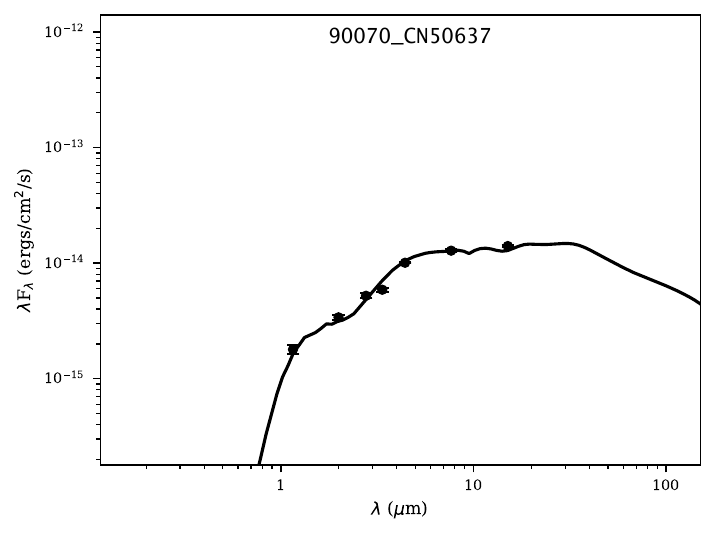}
\includegraphics{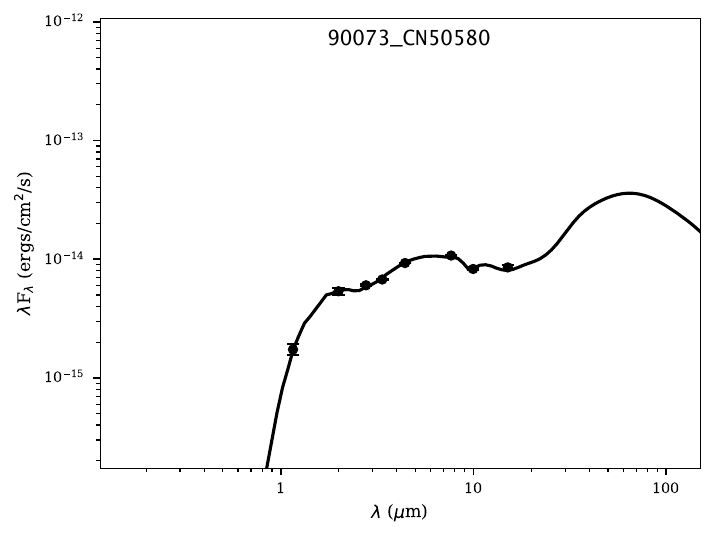}}
\resizebox{\hsize}{!}{
\includegraphics{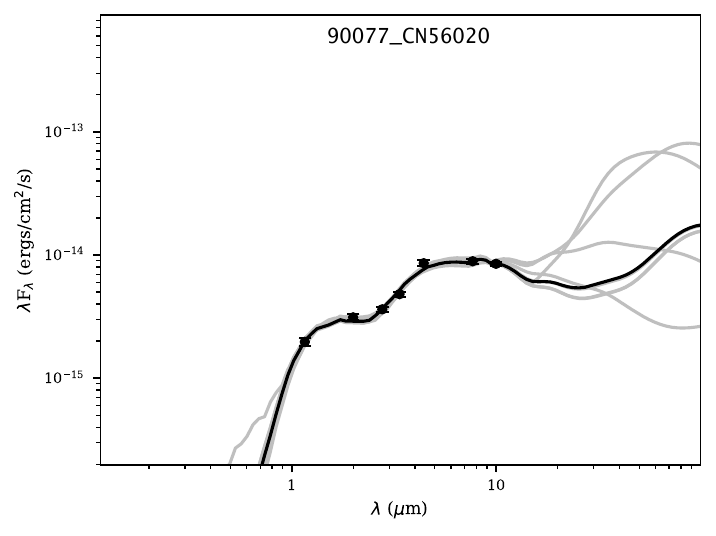}
\includegraphics{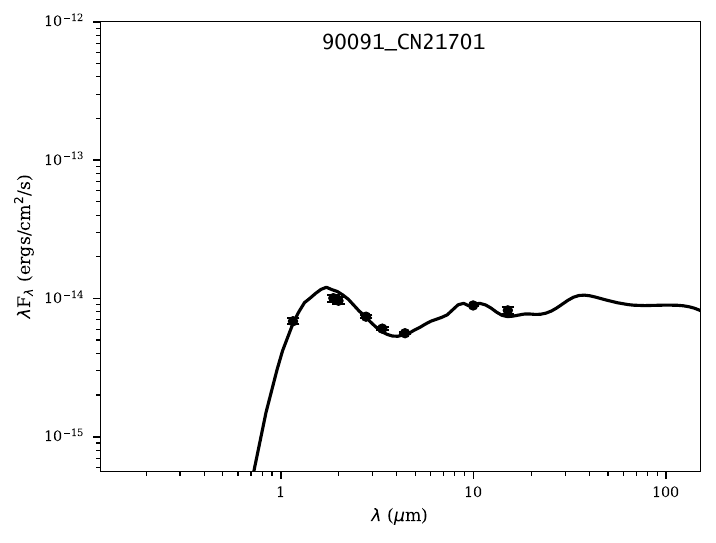}
\includegraphics{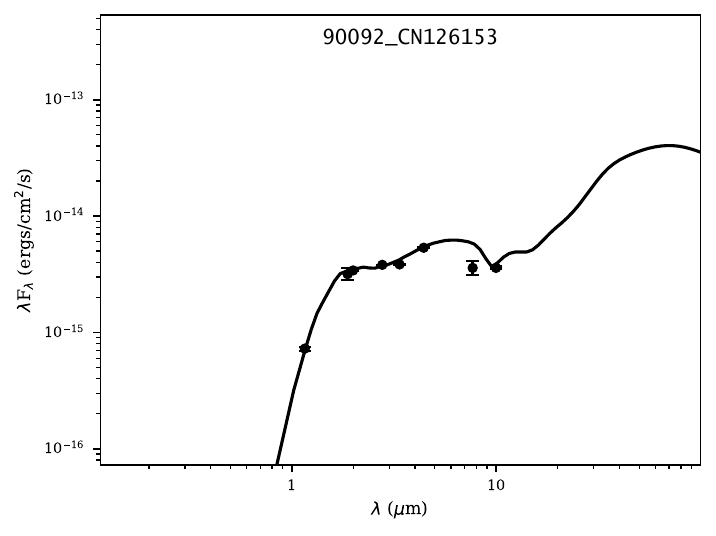}}
\resizebox{\hsize}{!}{
\includegraphics{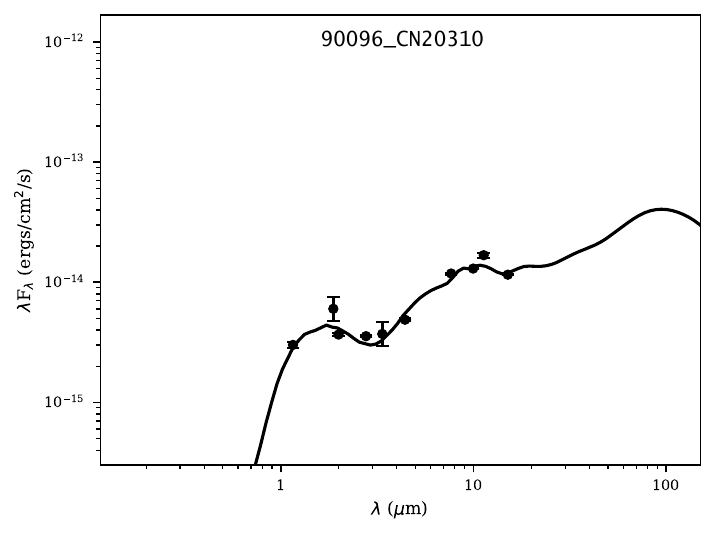}
\includegraphics{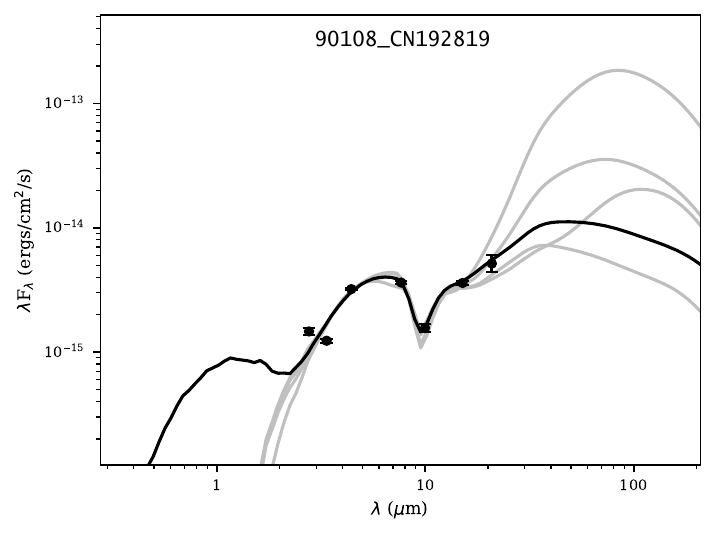}
\includegraphics{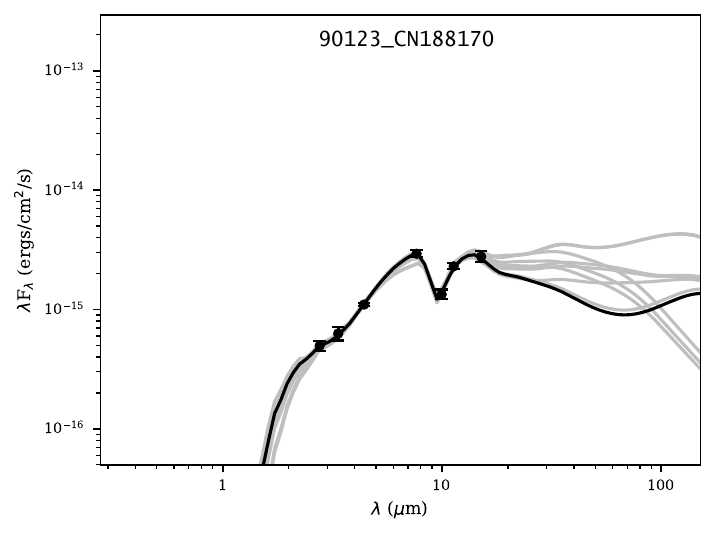}}
\caption{Best fit to the SEDs of the observed source, obtained using the models by \cite{robitailleetal2006,robitaille2017}. {Black lines correspond to the best-fit YSO model, while grey lines indicate fits with slightly larger $\chi^2$ but still within 10\,\% of the best fit (see \citealt{robitaille2017} for more details).} The best fitting parameters are shown in Tables\,\ref{tab1} and \ref{tab2}. }
\label{fig1}
\end{figure*}

The YSOs targeted with the NIRSpec MSA were selected from a list of 833 candidate YSOs that \cite{habeletal2024} drew from their combined near- and mid-IR photometric catalogue of NGC\,346 and identified via mid-IR excess. This set of sources was further supplemented with 182 YSOs identified in the same work by their near-IR excess. Spectral energy distributions (SEDs) for the sources in this combined list were then constructed from up to eleven near- and mid-IR photometric measurements. These SEDs were manually inspected for a rising profile toward the mid-IR and fitted to model YSO SEDs \citep{robitailleetal2006, robitaille2017, richardsonetal2024}, in order to further constrain their YSO nature and infer their stellar properties (including mass, radius, luminosity, and temperature). We then ordered the candidates by near-IR luminosity and gave the highest ranking to the brightest sources with the most photometric points, the most steeply rising SEDs, and were in general consistent with being YSOs based on our model fitting. We then manually inspected the highest ranked sources and discarded those with closely adjacent companions in near- and mid-IR imaging that would contaminate the MSA slit, or those with a diffuse morphology consistent with a background galaxy. Finally, we consulted the JWST exposure time calculator to establish a range of acceptable source luminosities that would produce an adequate signal-to-noise ratio ($\ga 10$) in the expected ice absorption features, without saturating in the $1.7 - 3.2$\,{\textmu}\,m wavelength range. We used as representative the Galactic ice-bearing YSO W33A \citep{gibbetal2000}, {having adopted a distance of $3.8$\,kpc \citep{immeretal2013} and} properly rescaled its luminosity to account for the 61\,kpc distance to NGC\,346  \citep{hilditchetal2005, kellerwood2006}. The final list of high-priority YSOs included approximately 150 sources. 

Using the eMPT software tool \citep{bonaventuraetal2023}, we identified the most suitable targets (and corresponding MSA configuration) for the specific orientation of the telescope at the time of the observations, avoiding sources that would have contaminants in nearby shutters for the selected orientation. We gave our YSO candidates the highest weight, and supplemented the list with PMS stars found by \cite{demarchietal2011} as well as main sequence stars, with a lower weight. With the selected MSA configuration, we were able to simultaneously observe 15 of our YSO sources, in addition to a number of PMS and main sequence (MS) sources that will be the subject of a future work (see Section 4.3). We show in Figure\,\ref{fig1} the fitted SEDs of the 15 YSOs. Most of them have spectral slopes consistent with Class I or earlier, although three sources ({90013}, 90069, and 90091) have a relatively flat profile that might in principle suggest a later evolutionary stage. The spectra of the 15 YSOs are discussed in the following sections.



\section{Observations and data reduction}
\label{observations}

The YSOs in our sample were observed with NIRSpec on 2024 April 30 using the G235M/F170LP and G395M/F290LP medium-resolution gratings ($R\sim 1000$) covering the range $1.7-5.3$\,{\textmu}m (proposal number 1227). A total of 104 sources were observed through the NIRSpec MSA, including the 15 YSOs, 36 PMS stars, and 53 main-sequence objects, using three different MSA configurations. In this work, we only discuss the spectra of the 15 YSOs, all of which were observed with both gratings. 

The MSA contains about 250,000 microshutters, each one defining a slit with a width of $0\farcs20$ and a length of $0\farcs46$, while the pitch between microshutters is $0\farcs27$ in the spectral direction and $0\farcs53$ in the spatial direction \citep{ferruit2022}. To properly measure (and subtract) the nebular background around our targets, we reserved for each source a ``slitlet'' composed of three neighboring microshutters in the cross-dispersion direction, spanning $0\farcs2 \times 1\farcs5$ on the sky.  

Each source was observed three times at each of the three different nodding positions in the slitlet, by repeatedly slewing the telescope in the cross-dispersion direction by an amount corresponding to the microshutter pitch ($0\farcs53$). In this way, each YSO was observed for a total of 5,383 s through each of the two gratings. 

The exposures were processed with the NIRSpec ramp-to-slope pipeline \citep{birkmannetal2022}, which performs bias subtraction, reference pixel subtraction, linearity correction, dark subtraction, and count-rate estimation, including jump detection and cosmic-ray rejection. From the resulting count-rate images, the flux- and wavelength-calibrated spectra were obtained using the NIRSpec Instrument Pipeline Software \citep{nips2018,alvesetal2018} to subtract the background, extract a subimage containing the spectral trace, assign wavelengths and spatial coordinates to all pixels, apply flat-fielding, and finally calibrate the flux. The resulting rectified spectra, resampled on a regular two-dimensional (2D) grid, correspond to arrays of $7 \times 1414$ pixel$^2$ for the F170LP/G235M grating, and $7 \times 1337$ pixel$^2$ for the F290LP/G395M combination, with the former covering the range $1.66 - 3.16$\,{\textmu}m and the latter $2.87 - 5.27$\,{\textmu}m, and a spatial extent of $0\farcs46$ in the cross-dispersion direction. 

Following \cite{demarchietal2024}, the one-dimensional spectrum of each source was computed by coadding the three pixel rows of the 2D-rectified product around the nominal source position, in order to optimise the signal-to-noise ratio (SNR). Although the target YSOs are the brightest sources in the slitlets at wavelengths longer than 3\,{\textmu}m, we nonetheless subtracted from their spectra the average spectrum extracted in the same way from the two neighboring microshutters, in order to remove the signature of the surrounding nebular background. 

{To account for possible local variations in the intensity of the nebular spectrum, following \cite{demarchietal2024} and \cite{rogersetal2025} the average background spectrum was rescaled so that the intensity of the He\,I emission line at $1.87$\,{\textmu}m matched the one observed in the source spectrum. The He\,I lines are fully nebular in nature and are not expected to be associated with the photosphere of our sources.} {The continuum level was determined through a cubic spline fit after having selected reference points on the spectrum away from any known absorption or emission features (see Figure\,\ref{fig5} in the Appendix).}

\begin{figure*}[t]
\centering
\resizebox{\hsize}{!}{\includegraphics[trim=100 50 50 50, clip]{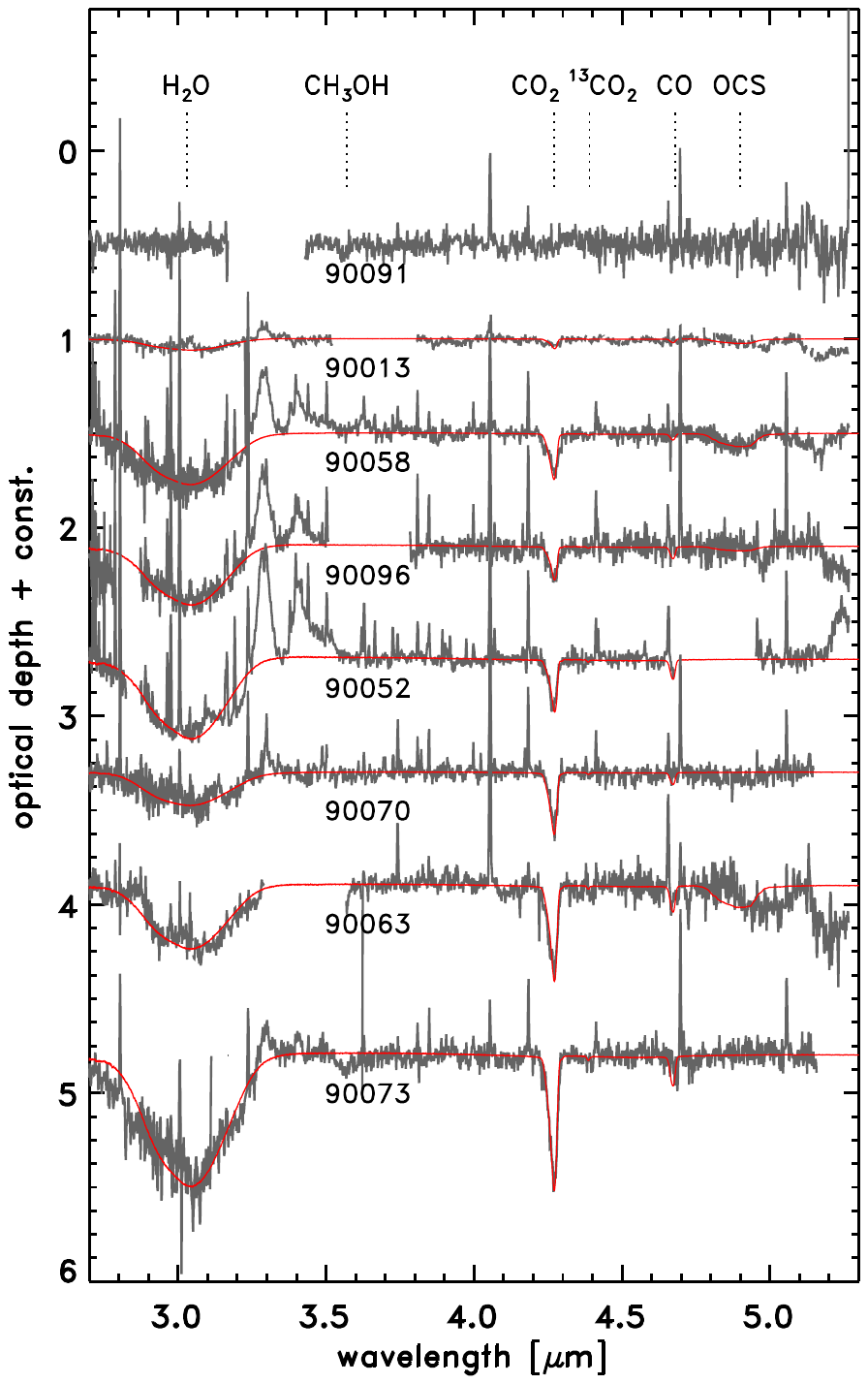}\includegraphics[trim=100 50 50 50, clip]{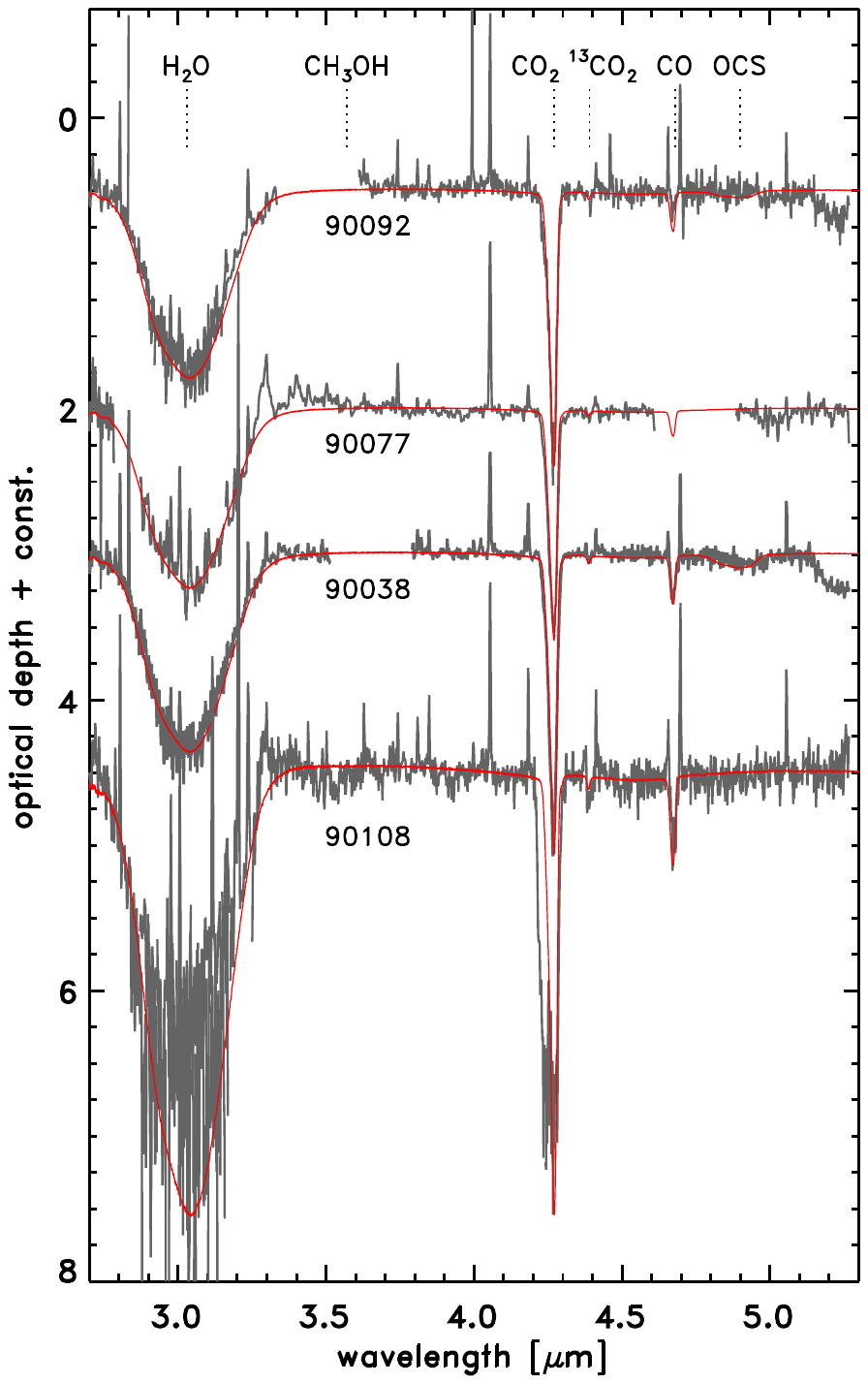}}
\caption{Optical depth spectra of the YSOs in our sample, shifted vertically for clarity. For illustration purposes, the red lines correspond to a linear combination of four ice mixtures selected from LIDA (3000 monolayers of pure H$_2$O at a temperature of 15\,K, H$_2$O:CO$_2$ in the ratio 10:1 at 10\,K, CO:CO$_2$ in the ratio 2:1 at 15\,K, and pure OCS at 17\,K) and provide a reasonable fit to the observations. The corresponding column densities are listed in Table\,\ref{tab4}. {The legend at the top of each panel indicates the ice features discussed in the text and the corresponding wavelengths, to guide the eye.}}
\label{fig2}
\end{figure*}

{Intermediate-mass YSOs like those studied in this work are known to exhibit gas-phase CO absorption near $4.7$\,{\textmu}m \citep[see, e.g.,][]{thietal2010}, which influences the continuum placement around the CO ice absorption band at $4.67$\,{\textmu}m. While the $R\sim1000$ resolution of our observations does not allow us to study these features in detail, the spectra do present enhanced fluctuations in the wavelength range expected for the P and R branches of CO gas, corresponding to changes in the rotational quantum number $\Delta J=-1$ and $\Delta J=+1$, respectively. In a conservative approach, we have placed the continuum at half-way intensity through those fluctuations.}

\section{Results}
\label{sec:results}

The resulting optical-depth spectra are shown in Figure\,\ref{fig2}. {The expected wavelengths of the ice features discussed here are indicated in the legend at the top of the figure, to guide the eye.} In four of the sources (90028, 90069, 90091, 90123), all ice features are absent or indistinguishable from the noise, even though the SEDs of two of them (90028, 90123) appear consistent with Class I sources. For this reason, we will not consider these four objects in the ice analysis that follows, although we show the spectrum of source 90091 in Figure\,\ref{fig2} for comparison.

Water and CO$_2$ absorption bands at $3.05$\,{\textmu}m and $4.27$\,{\textmu}m, respectively, are detected in all other eleven sources. Two of the sources, 90038 and 90108, also show clearly the $^{13}$CO$_2$ isotopologue band at $4.39$\,{\textmu}m, which is detected with a weaker optical depth also for sources 90052, 90077, and 90092. 

\begin{figure*}[t] 
\centering
\resizebox{\hsize}{!}{\includegraphics[trim=50 0 15 200, clip]{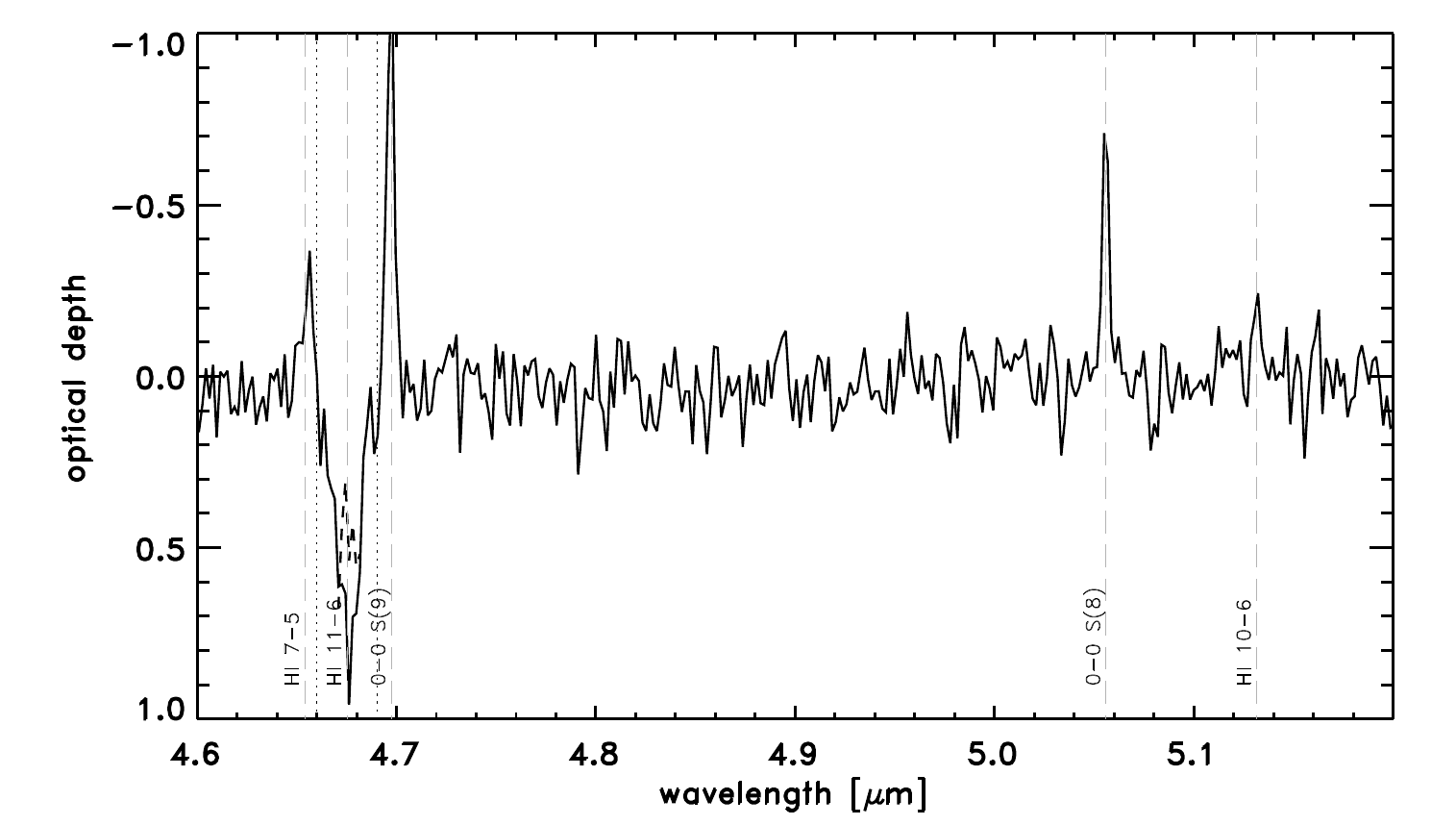}}
\caption{Optical depth spectrum of source 90108 before (dashed line) and after (solid line) subtraction of the Hu 11--6 recombination line. Vertical thin dashed lines mark the location of other hydrogen recombination lines and molecular hydrogen excitation lines, as indicated. Dotted lines show the column density integration range.}
\label{fig3}
\end{figure*}

As for the CO ice band at $4.67$\,{\textmu}m, it can be studied in all our sources but one, 90077, for which the relevant wavelength range falls in the gap between the two NIRSpec detectors and is not recorded. This is partly the case also for source 90052, although wavelengths shortwards of $4.676$\,{\textmu}m are properly measured and allow us to set a firm lower limit to the corresponding CO column density towards this star. In all other sources the CO ice band is properly detected, although for four of them (90013, 90058, 90070, 90096) the core of the band is (partly) filled by the prominent Humphreys (Hu) 11--6 recombination line at $4.673$\,{\textmu}m, which we removed when computing the column densities (see Section\,\ref{coldens}). To our knowledge, no previous works have attempted to evaluate the effect of the Hu 11-6 emission line on the optical depth of the CO ice band. 

In addition, four sources (90013, 90038, 90058, 90063) show a broad absorption band at $\sim 4.9$\,{\textmu}m that is consistent with the $4.90$\,{\textmu}m C--O stretch of carbonyl sulfide (OCS) ice \citep{geballeetal1985,ferranteetal2008}, which is detected here for the first time in an extragalactic YSO. For source 90058, the band is peaked exactly at $4.90$\,{\textmu}m, but the peak is at $4.92${\textmu}m for source 90038, and at $4.95${\textmu}m for source 90063. It might also be present in source 90013 and 90096 at a wavelength of $4.97${\textmu}m and a different shape, so here the detection is tentative. The profile, width, and peak of the OCS ice band depend strongly on the temperature and composition of the ice and on whether the ice is irradiated or not \citep{ferranteetal2008}, and this is the likely cause of the observed differences.

\subsection{Column densities}
\label{coldens}

We measured column densities for the three main ice species (H$_2$O, CO$_2$, and CO) by integrating the optical depth spectra over fixed wavelength ranges, respectively $2.8–3.25$, $4.2-4.3$, and $4.66–4.69$\,{\textmu}m. The adopted band strengths are, respectively, $2.0 \times 10^{-16}$, $7.6 \times 10^{-17}$, and $1.1 \times 10^{-17}$ cm\,molecule$^{-1}$ \citep{gerakinesetal1995,gibbetal2004}. The integration range for the H$_2$O ice avoids the PAH emission features seen in the spectra of Figure\,\ref{fig2} and the integration was applied to a heavily smoothed version of the spectra to avoid the effect of the many atomic hydrogen (H) and molecular hydrogen (H$_2$) emission lines. The integration range for the CO ice does specifically exclude the Pfund $\beta$ line at $4.65$\,{\textmu}m on one side and the molecular hydrogen S(9) excitation line at $4.697$\,{\textmu}m on the other, but not the Hu 11--6 recombination line at the centre of the band. 

\begin{deluxetable*}{lccccccc}
\tiny
\tablecaption{Column densities. \label{tab2}}
\tablehead{ \colhead{ID} & \colhead{Mass} & \colhead{$A_V$} & \colhead{$m_{227}$} & \colhead{$N$(H$_2$O)} & \colhead{$N$(CO$_2$)} & \colhead{$N$(CO)} & \colhead{$N(^{13}$CO$_2$)}\\
\colhead{} & \colhead{[$\mathrm{M_{\odot}}$]} & \colhead{}  & \colhead{} & \colhead{[cm$^{-2}$]} & \colhead{[cm$^{-2}$]} &  \colhead{[cm$^{-2}$]} &  \colhead{[cm$^{-2}$]} }
\decimals
\startdata
\hline
90013~~~~~~ & ~~4.1~~~ &  ~~3.4~~~ & ~~~17.1~~~ & ~$(6.9\pm1.5)\times10^{16}$~~ & ~$(1.4\pm0.3)\times10^{16}$~~ & ~$(2.9\pm0.6)\times10^{16}$~~ & --- \\
90038 & 3.0 &  9.8 & 19.0 & $(2.3\pm0.1)\times10^{18}$ & $(4.7\pm0.2)\times10^{17}$ & $(2.4\pm0.1)\times10^{17}$ & ~~$(3.2\pm0.2)\times10^{15}$~~\\
90052 & 3.1 & 11.7 & 18.8 & $(7.2\pm0.4)\times10^{17}$ & $(5.7\pm0.3)\times10^{16}$ & $(2.8\pm0.1)\times10^{16}$ & --- \\
90058 & 2.1 & 13.7 & 19.6 & $(6.1\pm0.3)\times10^{17}$ & $(6.2\pm0.3)\times10^{16}$ & $(2.9\pm0.9)\times10^{16}$ & $(1.7\pm0.1)\times10^{15}$ \\
90063 & 3.1 & 10.9 & 18.8 & $(4.0\pm0.2)\times10^{17}$ & $(1.2\pm0.1)\times10^{17}$ & $(8.7\pm0.4)\times10^{16}$ & --- \\
90070 & 2.2 &  9.3 & 19.9 & $(3.0\pm0.2)\times10^{17}$ & $(5.5\pm0.3)\times10^{16}$ & $(3.9\pm0.8)\times10^{16}$ & --- \\
90073 & 3.0 & 12.1 & 19.7 & $(1.2\pm0.1)\times10^{18}$ & $(1.6\pm0.1)\times10^{17}$ & $(1.1\pm0.1)\times10^{17}$ & --- \\
90077 & 2.7 & 8.3  & 20.3 & $(2.1\pm0.1)\times10^{18}$ & $(4.1\pm0.2)\times10^{17}$ & ---                        & $(6.1\pm0.4)\times10^{15}$ \\
90092 & 2.3 &  8.6 & 20.2 & $(2.1\pm0.1)\times10^{18}$ & $(4.4\pm0.3)\times10^{17}$ & $(9.3\pm0.5)\times10^{16}$ & $(4.0\pm0.3)\times10^{15}$ \\
90096 & 3.4 &  8.1 & 20.3 & $(5.7\pm0.3)\times10^{17}$ & $(4.6\pm0.2)\times10^{16}$ & $(4.1\pm0.7)\times10^{16}$ & --- \\
90108 & 1.8 & 27.3 & 21.3 & $(6.3\pm1.8)\times10^{18}$ & $(1.1\pm0.2)\times10^{18}$ & $(5.1\pm0.8)\times10^{17}$ & $(1.3\pm0.2)\times10^{16}$ \\
\hline
\enddata
\tablecomments{The columns are as follows: (1) source ID; (2) mass; (3) extinction $A_V$; (4) {magnitude in the NIRCam F227W band}; (5) H$_2$O ice column density; (6) CO$_2$ ice column density; (7) CO ice column density; (8) $^{13}$CO$_2$ ice column density.} 
\end{deluxetable*}

However, we are able to determine and remove the contribution of this emission line because our spectra cover also the Hu 10--6, Hu 12--6, and Hu 13--6 lines, which allow us to interpolate the expected luminosity for the Hu 11--6 line itself, based on the relative strengths of these hydrogen lines, assuming standard case B recombination \citep{hummerstorey1987} with a temperature $T_e=10,000$\,K and an electron density $N_e=10^4$ (understandably, the true line luminosity may differ if other line emission mechanisms are important, such as collisional excitation). Since all these lines are unresolved, we measured the optical depth of the CO ice band after subtracting a synthetic Hu 11--6 emission line with the interpolated luminosity and a Gaussian profile with the $R \sim 1200$ resolution of NIRSpec at this wavelength \citep{birkmannetal2022}. As an example, in Figure\,\ref{fig3} we show the optical depth spectrum of source 90108 around the CO band before the subtraction (thick dashed line) and after (solid line). In this case, the suppression of the Hu 11--6 line is important and the optical depth becomes $\sim 25$\,\% larger.

The derived column densities are listed in Table\,\ref{tab2}. Uncertainties on the column density include the measured statistical uncertainty on the observed NIRSpec spectra at the wavelengths of the ice bands, the uncertainty on the luminosity of the neighboring Hu emission lines when they are used to remove the Hu 11--6 line from the CO band, and an uncertainty on the determination of the continuum, which we have assumed to be 5\,\% for all sources. The latter is the largest source of uncertainty for most objects, {but not for those where the Hu 11--6 emission line was subtracted. For the relevant objects, Table\,\ref{tab3} provides the estimated flux of the Hu 11-6 emission line, the name and flux of the other Hu emission line used as a reference, and the flux measured in the CO ice absorption band, both before and after subtracting the Hu 11-6 line. All fluxes are in units of mJy {\textmu}m. The resulting uncertainty on the line flux after the subtraction is of the order of 20\,\% and is included in the column density uncertainties listed in Table\,\ref{tab2}. }.

\begin{deluxetable*}{lcccccc}
\tablecaption{Fluxes in the CO band before and after subtraction of the Hu 11--6 line. \label{tab3}}
\tablecolumns{7}  
\tablewidth{\linewidth}  
\tablehead{\colhead{ID} & \colhead{Flux CO} & \colhead{Ref. Line} & \colhead{Flux Ref. Line} & \colhead{Flux Hu 11--6} & \colhead{Flux CO} & \colhead{Uncertainty}\\
\colhead{} & \colhead{[mJy\,{\textmu}m]} & \colhead{} & \colhead{[mJy\,{\textmu}m]} & \colhead{[mJy\,{\textmu}m]} & \colhead{[mJy\,{\textmu}m]} & \colhead{}}
\decimals
\startdata
\hline
90013~~~~~ & ~~~~~$0.50\pm0.025$~~~~~ & ~~~~~Hu 13--6~~~~~ & ~~~~~$0.27\pm0.12$~~~~~ & ~~~~~$0.45\pm0.21$~~~~~ & ~~~~~$0.95\pm0.21$~~~~~ & $21.6$\,\% \\
90058 & $0.075\pm0.004$ & Hu 10--6 & $0.16\pm0.07$ & $0.12\pm0.05$ & $0.19\pm0.05$ &  $25.9$\,\% \\
90070 & $0.035\pm0.002$ & Hu 10--6 & $0.17\pm0.04$ & $0.12\pm0.03$ & $0.16\pm0.03$ &  $20.7$\,\% \\
90096 & $0.025\pm0.002$ & Hu 10--6 & $0.31\pm0.05$ & $0.23\pm0.04$ & $0.26\pm0.04$ &  $15.9$\,\% \\
90108 & $0.29\pm0.01$ & Hu 10--6 & $0.06\pm0.03$ & $0.04\pm0.02$ & $0.32\pm0.03$ & $8.7$\,\%\\ 
\hline
\enddata
\tablecomments{The columns are as follows: (1) ID of the source ; (2) absorption flux in the CO band in the observed spectrum; (3) Humphreys line used as a reference to estimate the flux of the Hu 11--6 emission feature; (4) Flux in the reference line; (5) Estimated flux of the Hu 11--6 emission line; (6) absorption flux in the CO band after subtraction; (7) Resulting uncertainty on the CO band flux after subtraction. All fluxes are in units of mJy\,{\textmu}m and are indicated with their $1\,\sigma$ uncertainty.}
\end{deluxetable*}

We have also included in Table\,\ref{tab2} the column densities of the $^{13}$CO$_2$ ice, measured for the five sources in which it is detected. We integrated the optical depth spectra over the range $4.38 - 4.40$\,{\textmu}m and adopted the band strength value of $7.1 \times 10^{-17}$ cm\,molecule$^{-1}$ from \cite{gerakineshudson2015}. Assuming that both isotopologues originate from the same region of the circumstellar disc, we can compute the average column density ratio $N(^{12}$CO$_2)/N(^{13}$CO$_2)$, which amounts to $82\pm21$ and agrees well, within the uncertainties, with the value of 77 measured in the local interstellar medium \citep{wilson1999}. 

{The absorption feature seen at $\sim 3.55$\,{\textmu}m in the spectrum of source 90073 is consistent with methanol ice (CH$_3$OH). The feature might be present also for source 90108, but it is noisier. Adopting a band strength of $5.3\times10^{-18}$ cm\,molecule$^{-1}$ from \cite{hudginsetal1993} and integrating the optical deph spectrum in the range $3.52 - 3.59$\,{\textmu}m, we obtain a methanol column density of $5.1\times10^{17}$ for source 90073. This is the first direct detection of methanol ice in the SMC.}

\subsection{Comparison with laboratory measurements}
\label{labmeas}
  
As a consistency check and to better constrain the temperatures of the ices, we compare our observations with spectra of ice analogues measured in the laboratory. The red lines in Figure\,\ref{fig2} correspond to optical depth spectra taken from the ``Leiden Ice Database for Astrochemistry'' (LIDA; see \citealt{rochaetal2022}). For illustration purposes, {in Figure\,\ref{fig2}} we selected from LIDA four representative ice mixtures, namely 3000 monolayers of pure H$_2$O at a temperature of 15\,K \citep{obergetal2007}, H$_2$O:CO$_2$ in the ratio 10:1 at 10\,K \citep{ehrenfreundetal1999}, CO:CO$_2$ in the ratio 2:1 at 15\,K \citep{vanbroekhuizenetal2006}, and pure OCS at 17\,K \citep[see][]{rochaetal2022}, and combined them linearly by adjusting the column densities of the four mixtures\footnote[1]{{The LIDA laboratory measurements for CO$_2$ necessarily include both the $^{12}$CO$_2$ and $^{13}$CO$_2$ isotopologues. The red lines in Figure\,\ref{fig2} always indicate a more or less pronounced dip at the wavelength of the $^{13}$CO$_2$ ice absorption band, which scales with the optical depth of the $^{12}$CO$_2$ band.}} to produce the synthetic optical depth spectra shown in red in Figure\,\ref{fig2}. The curves provide a reasonable fit to the observations, with the column densities listed in Table\,\ref{tab2}. 

As noted by \cite{rochaetal2022}, the fits obtained with this approach are neither ideal nor unique, so they must be considered as indicative. For instance, the CO$_2$ and $^{13}$CO$_2$ bands observed in the spectrum of source 90108 are broader and more structured than the synthetic spectra, which might indicate that a combination of mixtures is present along the line of sight, some dominated by water (polar) and some dominated by CO$_2$ (non-polar). Source 90108 (the object of lowest mass in our sample) shows a clear double-peaked CO$_2$ band, with maxima at $4.24$ and $4.27$\,{\textmu}m (and a possible third maximum at $4.23$\,{\textmu}m), which are likely associated with the presence of both pure and annealed CO$_2$ ices \citep{vanbroekhuizenetal2006}. 

The limitation of this approach is that the column densities listed in Table\,\ref{tab4} are specific to the selected laboratory mixtures of ice analogues, and not to the actual (unknown) composition responsible for the observed ice features. At the same time, a detailed analysis of the shapes of the observed absorption features can lead to the determination of the physical and chemical properties of the ice mixtures and can clarify their processing history \citep[e.g.][]{boogertetal2015}. This is beyond the scope of the present paper, but we note that the column density values in Table\,\ref{tab4} are broadly consistent with those derived in Section\,\ref{coldens} by directly integrating the absorption bands as observed in the spectra, and this allows us to set constraints on the temperature.

{To set constraints on the temperature, we used laboratory measurements for three different temperature ranges, namely 15\,K (as shown in Figure\,\ref{fig2}), 45\,K, and $60-70$\,K, adopting the same temperature for the three main volatiles. We cannot probe the effects of small temperature variations around these values, because there are no laboratory measurements in the LIDA database (for instance, spectra for pure H$_2$O ice from \citealt{obergetal2007} are only available in increments of 30\,K), so some level of degeneracy between the actual temperatures and column density of the ices is unavoidable. However, we can rather confidently exclude temperatures in the range $60-70$\,K because at these temperatures the laboratory spectra show: 1. no CO ice, contrary to what we observe; 2. a CO$_2$ ice band profile with full width at half maximum (FWHM) about}

\begin{deluxetable*}{lcccc}[t]
\tiny
\tablecaption{Ice spectra selected from LIDA with their temperatures and column densities scaled to match approximately the observed spectra in Figure\,\ref{fig2}. \label{tab4}}
\tablehead{ \colhead{ID} & \colhead{Pure H$_2$O, 15\,K} & \colhead{H$_2$O:CO$_2$ (10:1), 10\,K} & \colhead{CO:CO$_2$ (2:1), 15\,K} & \colhead{OCS, 17\,K} \\
\colhead{} & \colhead{~~~\cite{obergetal2007}~~~}  & \colhead{~~~\cite{ehrenfreundetal1999}~~~} & \colhead{~~~\cite{vanbroekhuizenetal2006}~~~} & \colhead{~~~\cite{rochaetal2022}~~~} \\
\colhead{} & \colhead{[cm$^{-2}$]}  & \colhead{[cm$^{-2}$]} & \colhead{[cm$^{-2}$]} & \colhead{[cm$^{-2}$]} }
\decimals
\startdata
\hline
90013~~~ & $9.2\times10^{16}$ & ---                & $2.3\times10^{16}$ & ---                \\ 
90038 & $1.1\times10^{17}$ & $3.6\times10^{18}$ & $4.0\times10^{17}$ & $4.0\times10^{16}$ \\
90052 & $6.6\times10^{17}$ & ---                & $1.2\times10^{17}$ & ---                \\
90058 & $1.5\times10^{17}$ & $5.0\times10^{17}$ & $3.8\times10^{16}$ & $3.0\times10^{16}$ \\
90063 & $3.0\times10^{17}$ & $4.1\times10^{17}$ & $1.7\times10^{17}$ & $5.0\times10^{16}$ \\
90070 & ---                & $5.0\times10^{17}$ & $7.6\times10^{16}$ & ---                \\
90073 & $5.9\times10^{17}$ & $9.1\times10^{17}$ & $1.9\times10^{17}$ & ---                \\
90077 & ---                & $3.5\times10^{18}$ & $2.1\times10^{17}$ & ---                \\
90092 & ---                & $3.6\times10^{18}$ & $3.2\times10^{17}$ & ---                \\
90096 & $4.8\times10^{17}$ & $1.6\times10^{16}$ & $7.6\times10^{16}$ & ---                \\
90108 & $2.4\times10^{18}$ & $4.3\times10^{18}$ & $7.4\times10^{17}$ & ---                \\
\hline
\enddata
\end{deluxetable*}

\noindent {30\,\% narrower than observed, and characterised by a blue shoulder not present in the data; and 3. an asymmetric H$_2$O ice band, peaked at $3.07$\,{\textmu}m while the one that we observe is more symmetrical and peaked at $3.04$\,{\textmu}m. Our observations also disfavour a temperature of 45\,K because of the asymmetry of the H$_2$O ice band and a 30\,\% lower FWHM of the CO$_2$ ice absorption band. This allows us to conclude that the observed ice bands indicatively agree with molecules at a temperature of approximately 15\,K.}

{However, there are clear indications that at least some of the ices have been exposed to warmer environments. The ``V'' shape of the water absorption band for sources 90073, 90077, 90092 (and possibly 90038) is a characteristic feature of crystalline H$_2$O ice \citep{smithetal1989}, as opposed to the broader, rounded, ``U''-shaped profile of amorphous ice present in the other sources.  Detection of crystalline ice indicates that the ice has been thermally processed. It could have been heated to temperatures above $\sim 100-110$\,K, even briefly, for instance in the warmer inner regions of the disc \citep{jenniskensblake1994}; or annealed through exposure to intermediate temperatures ($\sim 80-90$\,K) for a long period of time ($10^4 - 10^6$\,yr), for instance in the outer disk or protostellar envelope, especially in regions with modest heating from the central star \citep{kouchietal1994, vandishoeckblake1998, boogertetal2015}. Although the spatial resolution of our observations does not allow us to probe which is the dominant effect, this is the first time that the consequences of thermal processes are detected in a circumstellar disc outside the Milky Way.}

\subsection{Other spectral features}
\label{otherfeat}
 
The spectra shown in Figure\,\ref{fig2} are very rich. Besides the ice absorption bands, a number of emission features are present. This includes H recombination lines from the Paschen, Brackett, Pfund, and Humphreys series, some of which are visible in Figure\,\ref{fig2}. A large number of H$_2$ excitation lines are also present in the spectra. Since the nebular background has been subtracted from the spectra (see Section 3), the H and H$_2$ emission features are intrinsic to the sources and confirm their YSO nature. In a recent work \citep{demarchietal2024}, we studied a sample of $1-2$\,\Msolar\, PMS stars in this same NGC\,346 cluster, finding {evidence} that in this low-metallicity environment protoplanetary discs have longer lifetimes than in the solar neighborhood. Stars with estimated ages between 20 and 30 Myr still show strong accretion ($\sim 10^{-8}$\,\Msolar\,yr$^{-1}$) from the H recombination lines, and measured H$_2$ line ratios point to shocks in the discs, presumably from outflows, as the primary source of excitation, suggesting that the circumstellar discs have not yet been dispersed. In a forthcoming paper (De Marchi et al., in prep.), we will present the accretion properties and shock measurements of the more massive YSOs discussed here. 

In that work, we will also address the $3.3$\,{\textmu}m emission feature due to the aromatic C--H stretching mode of the PAH molecules, which is detected in Figure\,\ref{fig2} for all nine sources whose spectra fully cover that wavelength range. The feature is present also in source 90063 that covers those wavelengths only in part, while for the remaining source, 90092, the range $3.25 - 3.6$\,{\textmu}m falls in the gap between the NIRSpec detectors. {Furthermore, six of the sources (90052, 90058,  90073, 90077, 90096, and 90108) also show an aliphatic component at $3.4$\,{\textmu}m \citep[see, e.g., ][and references therein]{yangli2023}. The intensity ratio between the $3.4$\,{\textmu}m and $3.3$\,{\textmu}m  features of the six objects is $0.67$ on average, with a standard deviation of $0.15$. Using the relationships recently published by \citet[see equations 11 -- 13]{yangli2023}, for the typical temperatures of our objects in the range $6000-12000$\,K we derive an aliphatic fraction $\eta_{\rm ali} \simeq 9.2$\,\%, which is about 60\,\% larger than those measured in massive hotter Galactic YSOs \citep{yangli2023}.} A larger aliphatic fraction is expected when the effective temperature of the star is lower because aliphatic sidegroups are easier to maintain attached to PAHs in UV-poor regions.

\begin{figure*}[t]
\centering
\resizebox{\hsize}{!}{
\includegraphics[trim=70 180 50 200, clip]{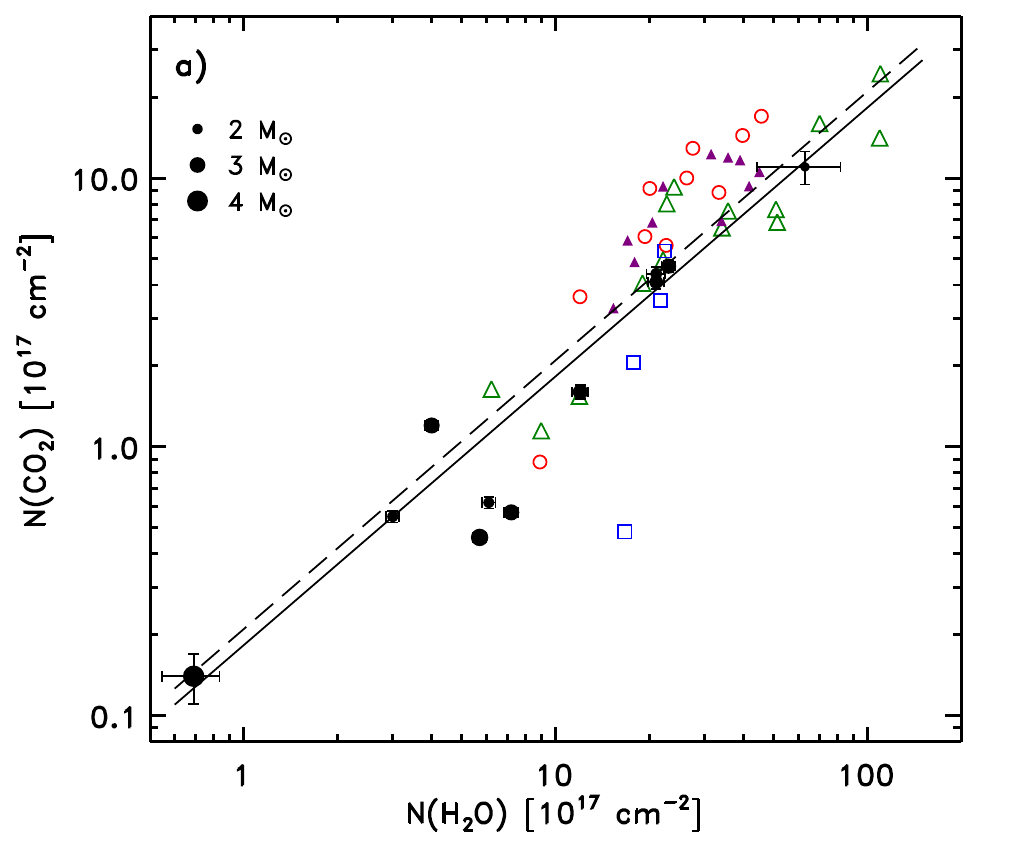}
\includegraphics[trim=70 180 50 200, clip]{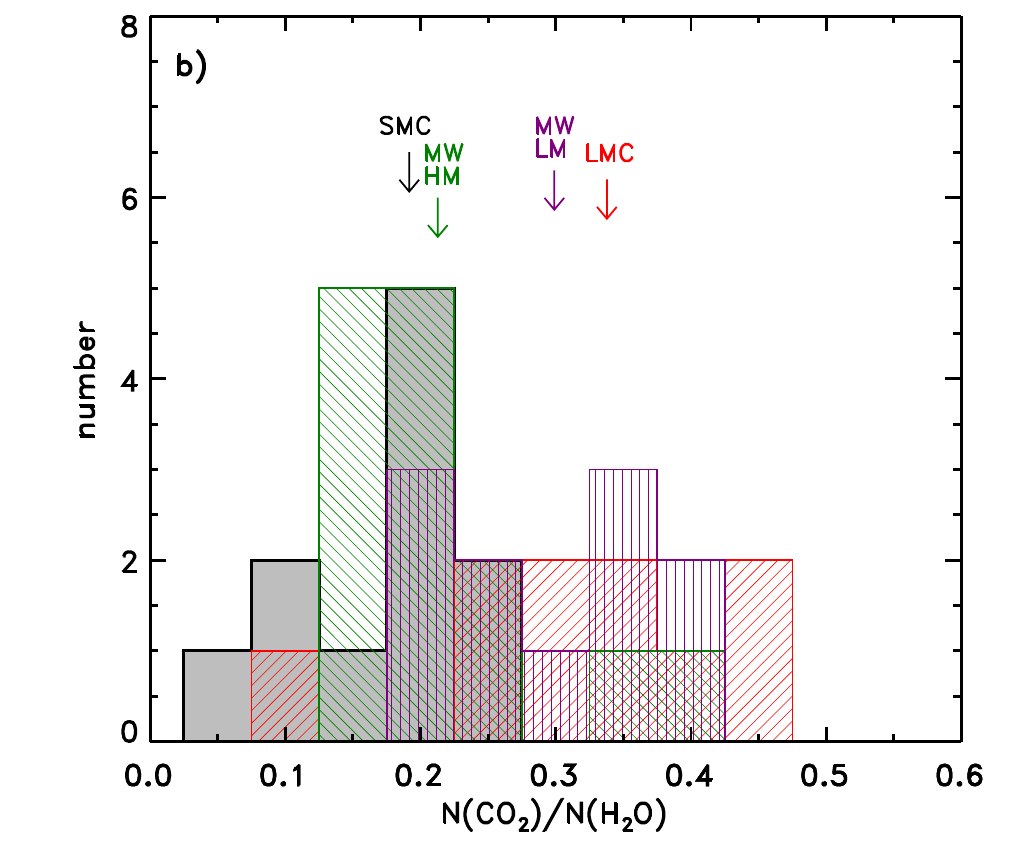}}
\resizebox{\hsize}{!}{
\includegraphics[trim=70 180 50 200, clip]{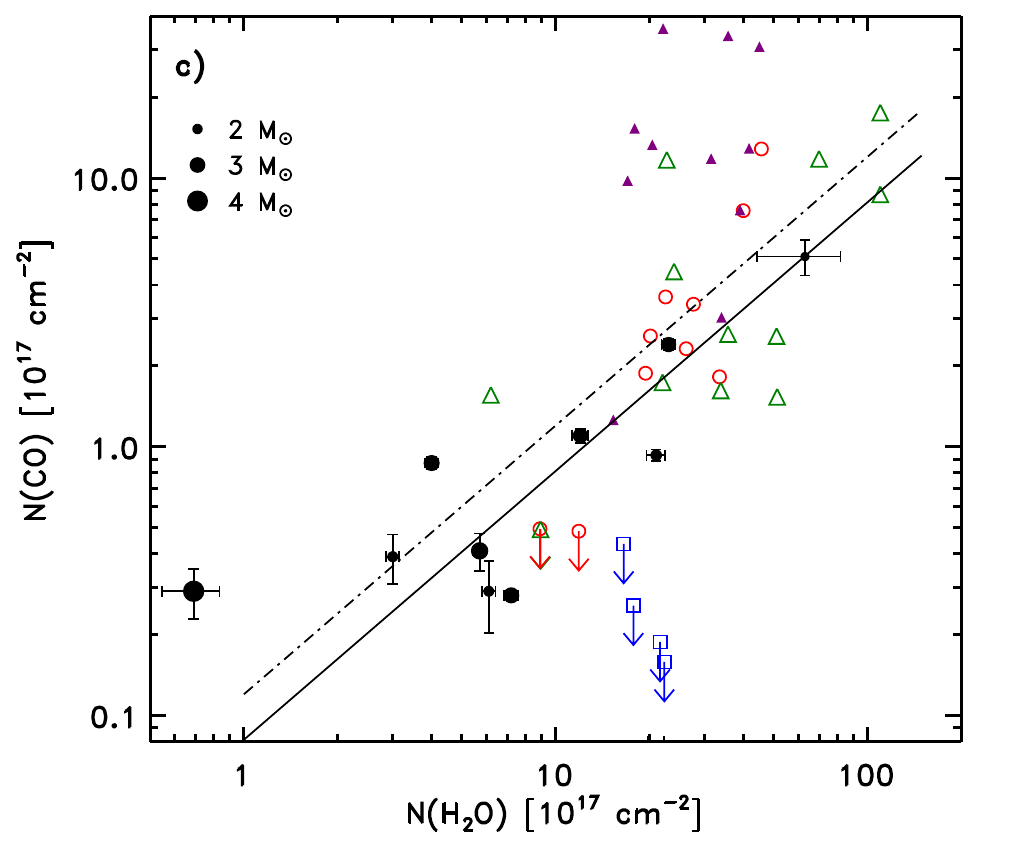}
\includegraphics[trim=70 180 50 200, clip]{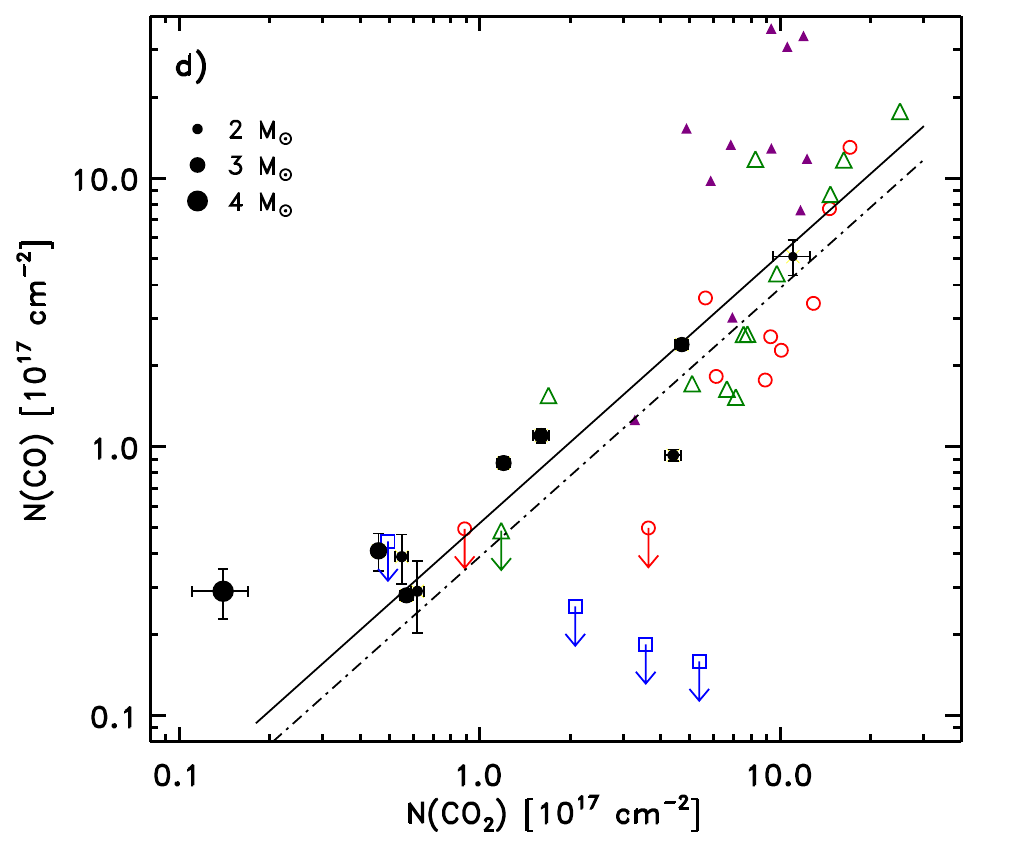}}
\caption{Panel {a)}: CO$_2$ vs H$_2$O ice column densities for the stars in NGC\,346 (black dots) compared with those measured for massive stars ($> 8$\,\Msolar) in the SMC (blue {squares}), in the LMC (red circles), and in the MW (large green triangles) as well as for low-mass Galactic YSOs (small purple triangles). The solid line is the best fit to the NGC\,346 column densities {and corresponds to a proportionality ratio of $0.18\pm0.06$}. The dashed line is the best fit to the ice column densities measured towards high-mass MW YSOs {and corresponds to a proportionality ratio of $0.21\pm0.08$}. Panel {b)}: Histograms showing the distribution of $N($CO$_2)/N($H$_2$O$)$ ice column density ratios in NGC\,346 (black line, gray shading), compared to the same in LMC massive protostars (red), in MW massive protostars (green), and in low-mass Galactic protostars (purple). Panels {c)} and {d)}: same as Panel a) but for $N($CO$)$ vs. $N($H$_2$O$)$ and $N($CO$)$ vs. $N($CO$_2)$, respectively. {The corresponding ratios, indicated by the solid lines, are $0.08\pm0.02$ and $0.52\pm0.16$, respectively. The dot-dashed lines show the mean ratios for the high-mass LMC stars alone.} Downward arrows are upper limits from the literature.}
\label{fig4}
\end{figure*}

\section{Discussion}
\label{discussion}

The column densities of the three major ice species are shown graphically in Figure\,\ref{fig4}, where they are compared to literature values. The column densities of the YSOs in NGC\,346 are shown as black dots, with a size proportional to the objects' mass, as per the legend. Also shown are the column densities measured for massive YSOs in the SMC (blue squares;  \citealt{oliveiraetal2011}), in the LMC (red dots; \citealt{shimonishietal2010,oliveiraetal2011}), in the Milky Way (large green triangles; \citealt{gerakinesetal1999,gibbetal2004}), as well as for low-mass Galactic YSOs (small purple triangles; \citealt{pontoppidanetal2003}). {Note that our $N($CO$_2)$ measurements come from the band at $4.23$\,{\textmu}m, while most of the literature values are obtained from the ground and come from the band at $15.6$\,{\textmu}m. This does not hinder the comparisons because our NIRSpec observations are not affected by the telluric lines that hamper ground-based measurements of the $4.23$\,{\textmu}m feature.}

The first difference that stands out clearly between NGC\,346 and the literature is that our YSOs, all of which have intermediate masses of $2-4$\,\Msun (see Table\,\ref{tab2}), extend the measurements to smaller column densities than those studied before: The ice abundances for all three species reach almost one order of magnitude lower than what was measured so far. 

In Figure\,\ref{fig4}a we show the column densities of H$_2$O and CO$_2$ ices. A weighted fit to our NGC\,346 points (black dots) passes through the origin, within the uncertainties. This confirms that direct proportionality between the two ice column densities (see \citealt{gerakinesetal1999}) applies also to intermediate-mass protostars in the SMC. The solid line in Figure\,\ref{fig4}a shows the best fit with a zero intercept and has a slope of $0.18\pm0.06$, corresponding to the $N($CO$_2)/N($H$_2$O$)$ ratio of column densities. This is in very good agreement with the value of $0.21\pm0.08$ compiled for massive Galactic YSOs by \cite{oliveiraetal2011}, based on previous measurements (large green triangles) by \cite{gerakinesetal1999} and \cite{gibbetal2004}, and shown by the dashed line in the same figure. The agreement between the column density ratios of these two samples (intermediate-mass stars in NGC\,346 and high-mass YSOs in the Galaxy) is also confirmed by the histogram shown in Figure\,\ref{fig4}b, where the two populations show very consistent distributions. {However, high-mass LMC YSOs and low-mass Galactic YSOs are known to deviate, with a systematically larger $N($CO$_2)/N($H$_2$O$)$ (see Section\,5.1).} 

{To test the direct proportionality between $N($CO$_2)$ and $N($H$_2$O$)$ we computed the ``coefficient of determination'' $R^2$, which indicates the proportion of 
the variance in a dependent variable $y$ that can be predicted from the independent variable $x$ in a regression model\footnote[3]{For a zero-intercept fit like ours, $R^2=1-\frac{\sum_i (y_i-ax_i)^2}{\sum_i y_i^2}$, with $a$ the slope.} \citep[see, e.g.,][]{iglesiasnavarroetal2024}.  $R^2$ is equal to 1 when all points fall exactly on the regression line. For the  $N($CO$_2)$ vs. $N($H$_2$O$)$ relationship we find $R^2=0.99$, indicating a very strong fit and confirming direct proportionality between the ice column densities.}

For our sources, a direct proportionality relationship (with zero intercept) provides the best fit also between the CO and H$_2$O ice column densities (Figure\,\ref{fig4}c) and between the CO and CO$_2$ ice column densities (Figure\,\ref{fig4}d). The solid line in Figure\,\ref{fig4}c corresponds to a ratio $N($CO$)/N($H$_2$O$)=0.08\pm 0.02$, {with a very high $R^2=0.93$ value, confirming again direct proportionality}. The line reproduces rather well also the distribution of ice abundances for high-mass YSOs in the MW and {reasonably well} those in the LMC, but low-mass Galactic YSOs (small triangles) show a larger abundance of CO ice when compared to water ice (see Section\,5.1). {The solid line in Figure\,\ref{fig4}d corresponds to a ratio $N($CO$)/N($CO$_2)=0.52\pm0.16$ for our NGC\,346 sources {(with $R^2=0.93$)} and agrees with the overall trend of the massive Galactic and LMC YSOs.} {Concerning the high-mass LMC YSOs, the small discrepancies in both Figure\,\ref{fig4}c and \ref{fig4}d are in fact consistent with the uncertainties. The mean $N($CO$)/N($H$_2$O$)$ abundance ratio of the LMC massive YSOs in the sample of \cite{oliveiraetal2011} is $0.12$ (shown by the dot-dashed line in Figure\,\ref{fig4}c) and within $2\,\sigma$ it agrees with our ratio of $0.08\pm0.02$. The same is true for Figure\,\ref{fig4}d, where the mean abundance ratio $N($CO$)/N($CO$_2)=0.39$ of the same LMC stars (dot-dashed line) agrees with our ratio of $0.52\pm0.16$ to within $1\,\sigma$. } 

\subsection{High-mass LMC YSOs and low-mass MW YSOs} 

As already pointed out by \cite{oliveiraetal2013}, the $N($CO$_2)/N($H$_2$O$)=0.34\pm0.09$ ratio measured for LMC high-mass YSOs by \cite{shimonishietal2010} and \cite{oliveiraetal2011} are systematically shifted to larger values {than in the MW and SMC} (Figure\,\ref{fig4}a and \ref{fig4}b), indicating a larger abundance of CO$_2$ ice compared to water ice. {In principle, this might reflect a different disc chemistry in the LMC, but if the differences were caused by metallicity one would expect even larger departures in the SMC, due to its even lower metal content. This is not what we and  \cite{oliveiraetal2011} observed.} 

{Currently there is no clear explanation for the apparent CO$_2$ column density excess in the LMC, but thanks to NIRSpec it is now possible to look for ices also around intermediate mass YSOs in that galaxy. An ideal place to investigate this issue further is 30\,Dor, where one can explore whether the CO$_2$ column density remains proportionately higher for YSOs across a wide range of masses.} {This raises an important question about the role of stellar mass in shaping ice abundances.}

The low-mass Galactic YSOs observed by \cite{pontoppidanetal2003} {also} occupy a different region in all panels of Figure\,\ref{fig4}, with a systematically larger proportion of CO$_2$ and CO ice compared to water. The mass of these objects is one order or magnitude smaller than that of our NGC\,346 sample and almost two orders of magnitude less than the massive YSOs shown in Figure\,\ref{fig4}. This could suggest that low mass might play a role in enhancing the CO and CO$_2$ column densities with respect to water. However, our column density ratios for objects with masses $2-4$\,\Msolar\,appear to be in line with those of the massive stars, rather than occupying an intermediate position. In other words, our data do not show an obvious transition to higher CO and CO$_2$ ice abundances with decreasing mass, at least not down to our lowest mass range of about 2\,\Msolar. It is possible that a transition phase might occur at lower masses. {It would be interesting to explore whether very low-mass stars have enhanced CO and CO$_2$ ice column densities also in low-metallicity environments. Suitable places to test this hypothesis are the star clusters in the outer Galaxy. With a metallicity $\sim 0.2$\,Z$_\odot$ and distances in the range $8-10$\,kpc, low-mass stars in these clusters can be studied efficiently with the NIRspec MSA.}

\subsection{No observed thresholds for CO or CO$_2$ ices}

{An important new result of this work is that we clearly detect for the first time the signature of CO ices in SMC YSOs}. As mentioned in the Introduction, \cite{oliveiraetal2011} studied four massive ($L>10^4$\,\Lsol) YSOs in the SMC using ground-based Very Large Telescope (VLT) spectroscopy and did not detect the $4.67$\,{\textmu}m absorption band towards them (hence the upper limits reported in Figure\,\ref{fig4}), although with the same instrumentation they did detect the presence of CO ices towards four equally massive YSOs in the LMC. They suggested that the lower metallicity and harsher environmental conditions in the SMC might be responsible for the observed lower gas-phase CO density \citep{leroyetal2007, jamesonetal2018}. 

Besides the lower metallicity, the SMC is also characterised by a particularly low C/H ratio of just 6\,\% compared to the solar value \citep{dufouretal1982,draine2003}. This lower gas-phase CO density, coupled with the higher dust temperature in SMC YSO envelopes \citep{vanloonetal2010}, might then inhibit the freeze-out of CO. Our detection of CO ice towards intermediate mass YSOs in NGC\,346 is not in line with this scenario, although one cannot exclude that the environment around the four high-mass YSOs studied by \cite{oliveiraetal2011} might be particularly harsh, more than in NGC\,346. However, a perhaps more likely possibility is that, at a distance of 62\,kpc, the CO ice absorption signature in the spectra of SMC YSOs is simply too weak to be detected from the ground, even with the VLT, due to the strong telluric CO contamination at $4.68$\,{\textmu}m, which is particularly difficult to subtract. For this reason, we will not consider the upper limits of \cite{oliveiraetal2011} in this analysis.

The $N($CO$)$ vs $N($CO$_2)$ relationship that we find in NGC\,346 is also very interesting. Previous studies of ice column densities towards Galactic and LMC massive protostars \citep{gerakinesetal1999,oliveiraetal2011, oliveiraetal2013} revealed no CO ice when $N($CO$_2)\la 4 \times 10^{17}$\,cm$^{-2}$, suggesting the presence of a detection threshold. In those studies, no secure detection of CO ices exists for $N($CO$)\la 10^{17}$\,cm$^{-2}$, only upper limits. But in our sample, CO ice is detected and its proportions relative to CO$_2$ are the same as for massive stars. 

{Our results also do not support previous claims of possible density thresholds for CO$_2$ ices. Based on the observation} of three massive YSOs in the SMC,\footnote[2]{The three sources are not included in Figure\,\ref{fig4} because no CO ice measurements were taken by \cite{oliveiraetal2013}.} with $N($H$_2$O$) \simeq (17.7 \pm 0.9) \times 10^{17}$\,cm$^{-2}$ and N$($CO$_2) \simeq (1.4 \pm 0.3) \times 10^{17}$\,cm$^{-2}$, \cite{oliveiraetal2013} suggested the presence of a column density threshold of $N($H$_2$O$)\simeq 1.5 \times 10^{18}$\,cm$^{-2}$ for the detection of CO$_2$. Our observations (Figure\,\ref{fig4}a) do not support this hypothesis.

\subsection{No metallicity dependence}

{Taken together, these findings} indicate that the lower gas metallicity in NGC\,346 does not prevent or inhibit the formation of CO ices, nor does it reduce the relative abundance of CO$_2$ ices, as suggested instead by \cite{oliveiraetal2011, oliveiraetal2013}. Indeed, while all three ice species have lower abundances in NGC\,346, their ratios are the same in the SMC as in the MW, as if the harsher radiation environment of the SMC were to affect in a similar way the three ice species, rather than preferentially the CO ice, which has the lowest sublimation temperature. Therefore, the main conclusion appears to be that the ice chemistry is the same in the SMC as in the MW, with no clear effect due to the lower metallicity and dust abundance and the lower shielding that they would imply. 

{A likely explanation is that} the ices we are detecting are in a portion of the disc that is well shielded, possibly in the protostellar envelope itself. Indeed, all our sources but 90013 have optical extinction $A_V > 8$ (see Table\,\ref{tab2}) and as such they comfortably exceed the $A_V=6$ threshold for the detection of CO ice in Galactic sources \citep{berginetal2005, whittetetal2007}, thereby making it possible for all three ice species to be present and for the temperature to remain around 15\,K. This is consistent with observations of heavily irradiated discs suggesting that the surface density of the disc provides sufficient shielding of the midplane, which remains mostly molecular in nature and is minimally affected by the environment \citep{walshetal2013, ramireztannusetal2023}. Furthermore, our YSOs are of intermediate mass and with a median temperature of $\sim 6600$\,K and median luminosity of $\sim 45$\,\Lsol\, their UV radiation field is lower than that typical of the MW and LMC high-mass objects reported in Figure\,\ref{fig4} \citep[see][]{gerakinesetal1999, gibbetal2004, oliveiraetal2011, oliveiraetal2013}.

\section{Conclusions}

{For the first time we have detected volatile ices in intermediate mass YSOs outside the MW. Our observations of NGC\,346 in the SMC provide a consistent picture of ice chemistry in low-metallicity environments. We find that, despite overall lower ice column densities, the relative abundances of H$_2$O, CO$_2$, and CO ices are similar to those observed in the Milky Way for massive stars. These ratios are maintained even at column densities lower than previously measured, with no evidence for the thresholds reported in earlier studies. The detection of CO ice, in particular, challenges earlier claims that low metallicity or harsh radiation fields in the SMC inhibit its formation. Instead, the observed trends suggest that the chemistry of the ices, including their formation and survival, is remarkably robust and not strongly affected by metallicity, at least within the mass range and environments probed by our sample.}

{These} results are compelling {because} all three ice species are observed simultaneously in the same spectrum, within a narrow wavelength range and using a single grating. {While} NGC\,346 may not be {representative} of the entire SMC, {our findings suggest that}, at least in this region, the chemistry of volatile ices is not significantly affected by low metallicity. If similar abundance ratios are confirmed in other low-metallicity star-forming regions, {for instance through future JWST observations}, {it would imply that the basic ingredients for forming planetary atmospheres, such as H$_2$O, CO$_2$, and CO}, were already available in comparable proportions early in cosmic history. This raises the possibility that planets with atmospheric compositions similar to those in the solar neighborhood could have formed as early as redshift~2 or higher.


\section{Acknowledgements}

We are very grateful to an anonymous referee for providing insightful comments that have helped us improve the presentation of our results. We thank Ewine van Dishoeck and Melissa McClure for useful discussions. This work is based on observations made with the NASA/ESA/CSA James Webb Space Telescope. All of the data presented in this paper were obtained from the Mikulski Archive for Space Telescopes (MAST) at the Space Telescope Science Institute, which is operated by the Association of Universities for Research in Astronomy, Inc., under NASA contract NAS 5-03127 for {\em JWST}. These observations are associated with program \#1227. The specific observations analyzed here can be accessed via~\dataset[10.17909/7bq5-j974]{https://doi.org/10.17909/7bq5-j974}.

NH and MM acknowledge that a portion of their research was carried out at the Jet Propulsion Laboratory, California Institute of Technology, under a contract with the National Aeronautics and Space Administration (80NM0018D0004). NH and MM acknowledge support through NASA/{\em JWST} grant 80NSSC22K0025. 
KB acknowledges support through Grant INAF 2022 TRAME@JWST (TRacing the Accretion Metallicity rElation-ship with NIRSpec@JWST).
ES is supported by the international Gemini Observatory, a program of NSF NOIRLab, which is managed by the Association of Universities for Research in Astronomy (AURA) under a cooperative agreement with the U.S. National Science Foundation, on behalf of the Gemini partnership of Argentina, Brazil, Canada, Chile, the Republic of Korea, and the United States of America.
OCJ acknowledges support from an STFC Webb fellowship. 
ON acknowledges the NASA Postdoctoral Program at NASA Goddard Space Flight Center, administered by Oak Ridge Associated Universities under contract with NASA.

\facilities{\em JWST} (NIRSpec)

\software{eMPT \cite{bonaventuraetal2023}}

\appendix
\renewcommand{\thefigure}{A\arabic{figure}} 
\setcounter{figure}{0}  

\begin{figure*}[ht]
\centering
\resizebox{\hsize}{!}{\includegraphics[trim=0 33 0 310, clip]{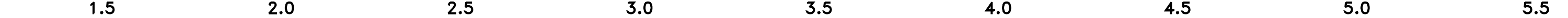}}
\resizebox{\hsize}{!}{\includegraphics[trim=0 50 0 28, clip]{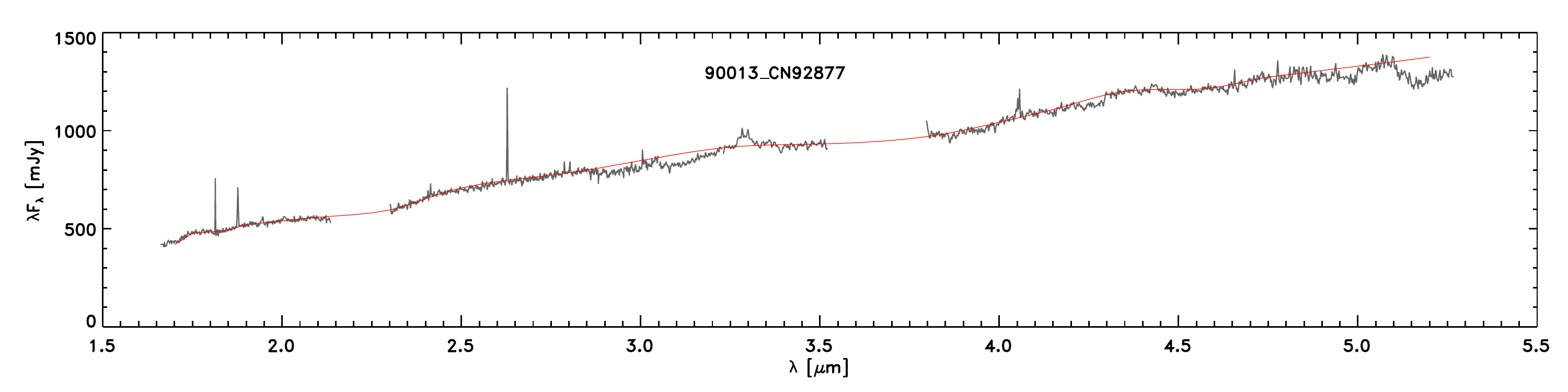}}
\resizebox{\hsize}{!}{\includegraphics[trim=0 50 0 28, clip]{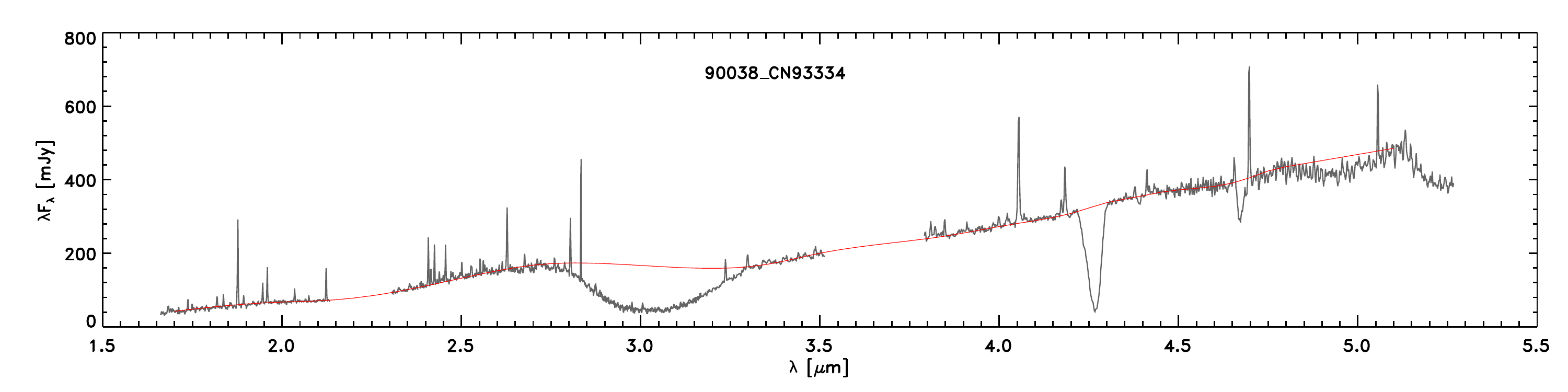}}
\resizebox{\hsize}{!}{\includegraphics[trim=0 50 0 28, clip]{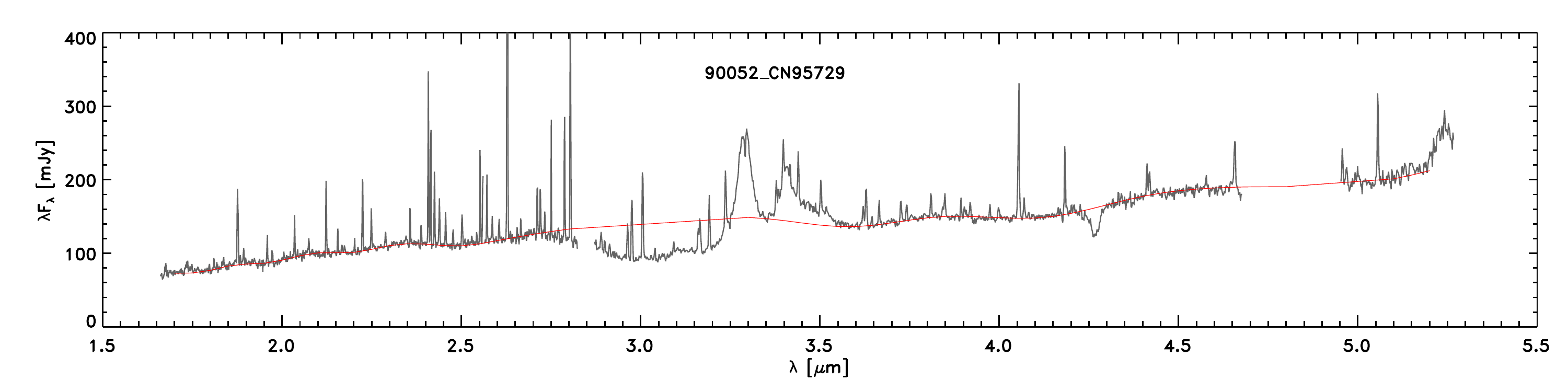}}
\resizebox{\hsize}{!}{\includegraphics[trim=0 50 0 28, clip]{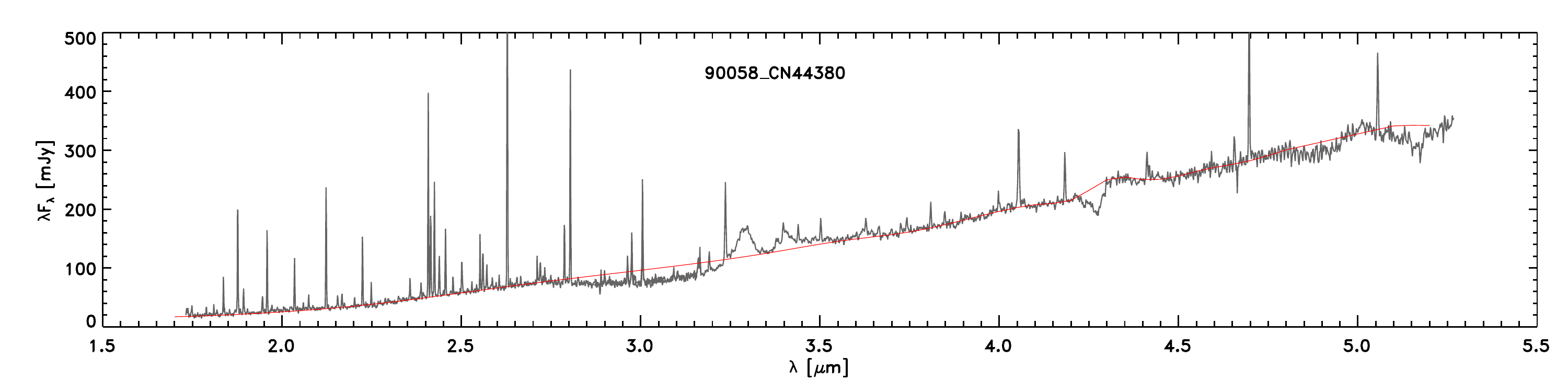}}
\resizebox{\hsize}{!}{\includegraphics[trim=0 50 0 28, clip]{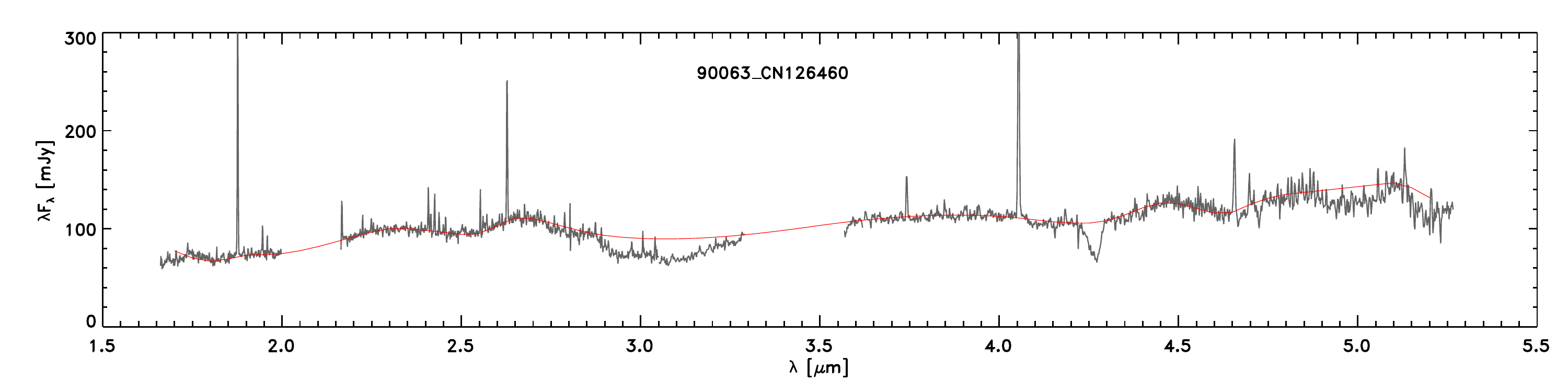}}
\resizebox{\hsize}{!}{\includegraphics[trim=0 10 0 28, clip]{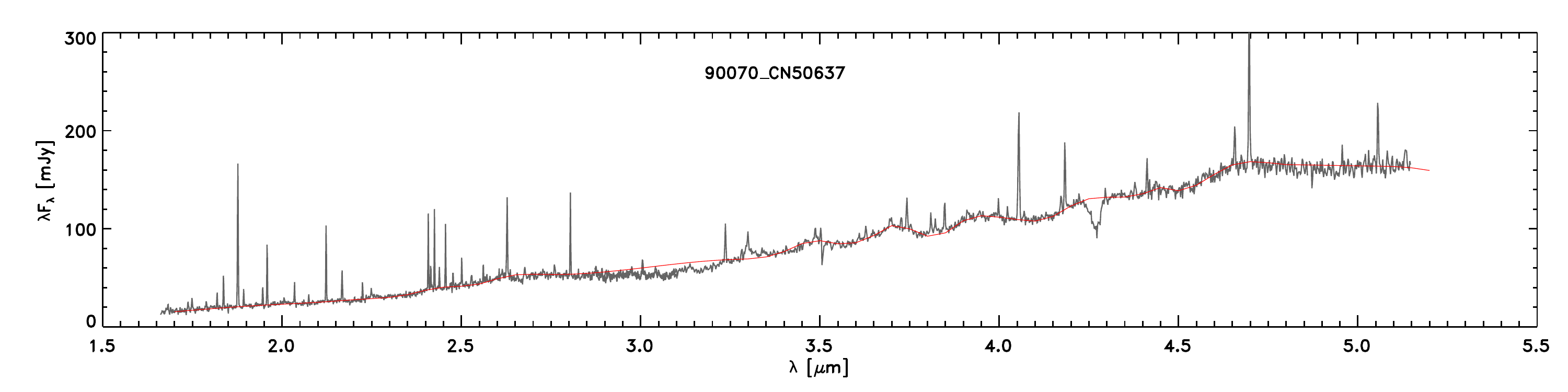}}
\caption{Observed spectra and fitted continua.}
\label{fig5}
\end{figure*}

\begin{figure*}[h]
\centering
\ContinuedFloat
\resizebox{\hsize}{!}{\includegraphics[trim=0 33 0 310, clip]{n346fig431wide.pdf}}
\resizebox{\hsize}{!}{\includegraphics[trim=0 50 0 28, clip]{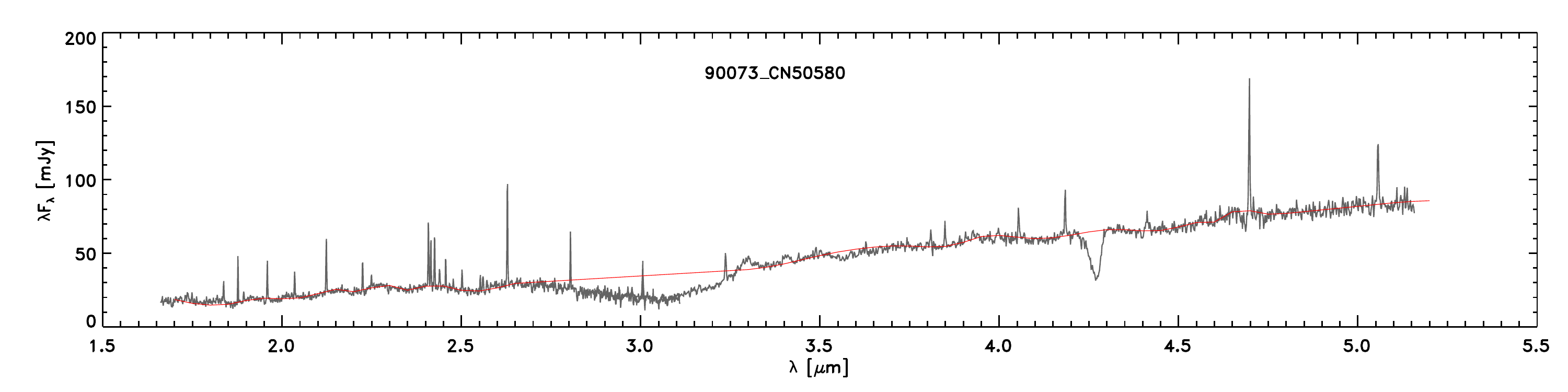}}
\resizebox{\hsize}{!}{\includegraphics[trim=0 50 0 28, clip]{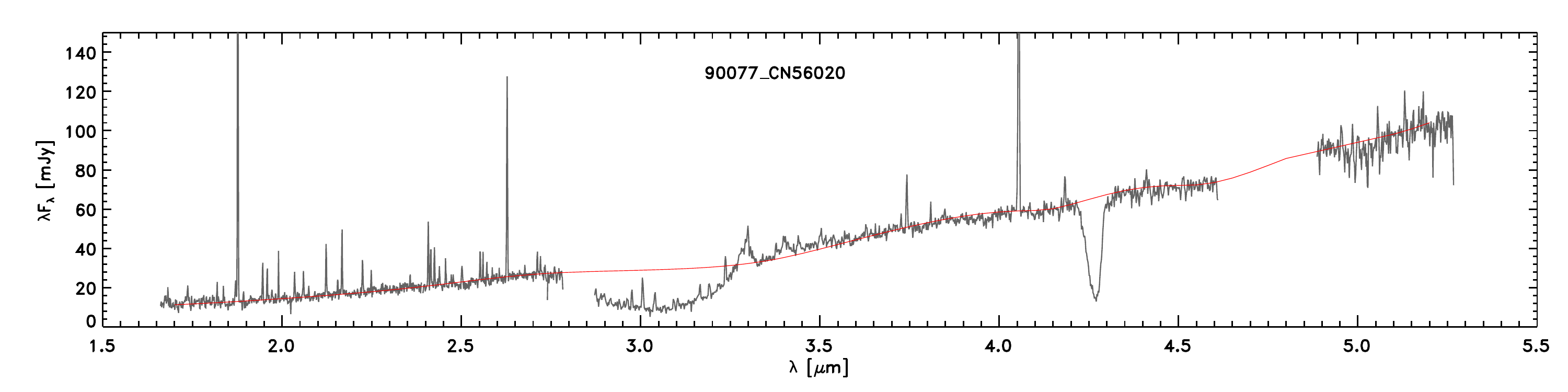}}
\resizebox{\hsize}{!}{\includegraphics[trim=0 50 0 28, clip]{n346fig431wide.pdf}}
\resizebox{\hsize}{!}{\includegraphics[trim=0 50 0 28, clip]{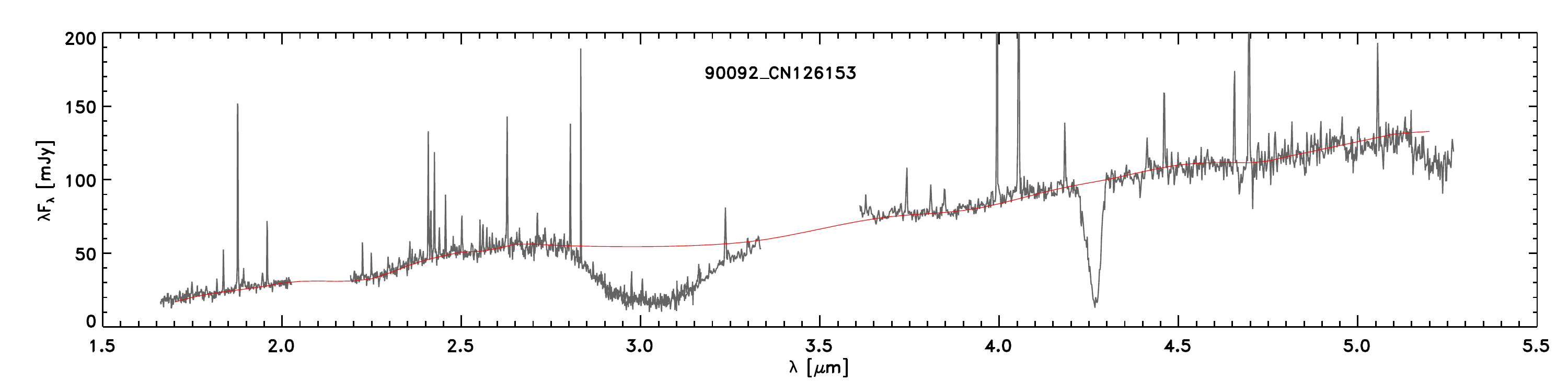}}
\resizebox{\hsize}{!}{\includegraphics[trim=0 50 0 28, clip]{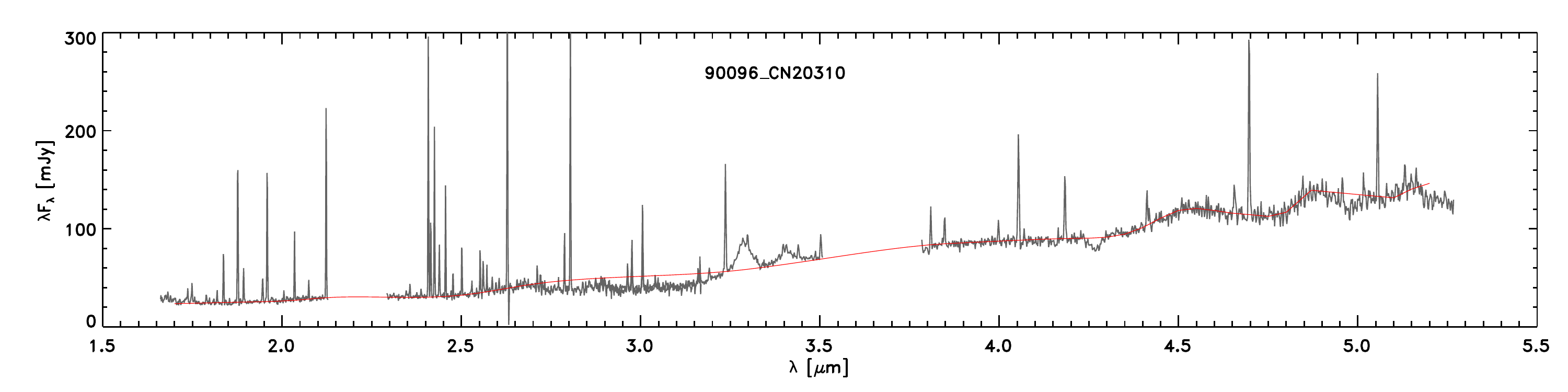}}
\resizebox{\hsize}{!}{\includegraphics[trim=0 10 0 28, clip]{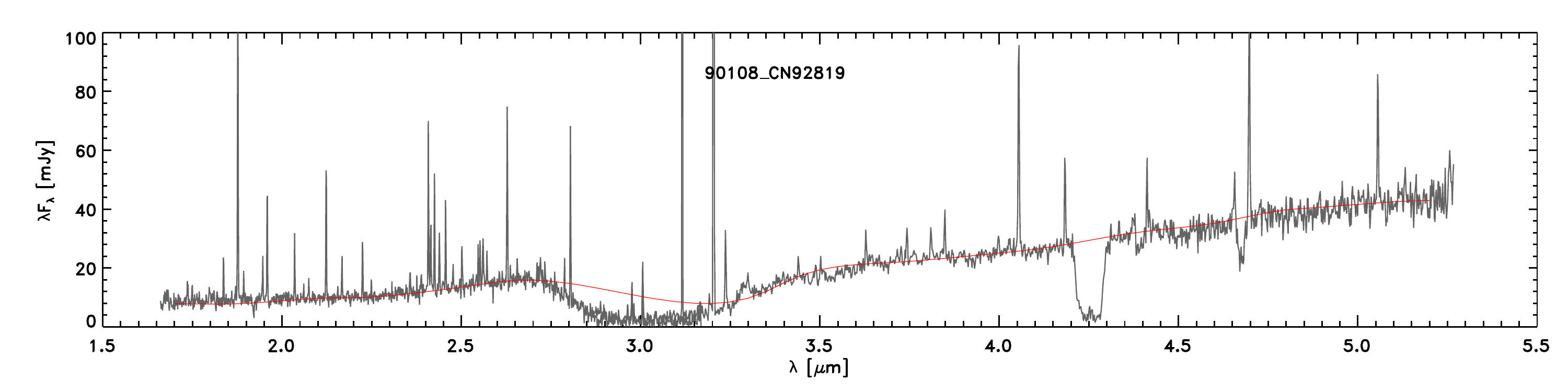}}
\caption{Continued.}
\label{fig5}
\end{figure*}

\FloatBarrier

\end{document}